\documentclass[aps,twocolumn,preprintnumbers,amsmath,amssymb,superscriptaddress]{revtex4-1}

\usepackage{graphicx}    
\usepackage{dcolumn}    
\usepackage{bm}             
\usepackage{color}
\usepackage{url}
\usepackage{wasysym}
\usepackage{natbib}
\usepackage{fixmath}

\begin{document}

\title{Raman spectroscopy of femtosecond multi-pulse irradiation \\ of vitreous silica: experiment and simulation}

\author{N.S. Shcheblanov}
\email{nikita.shcheblanov@polytechnique.edu}
\affiliation
{
Centre de Physique Th\'eorique, CNRS \& Laboratoire des Solides Irradi\'es CEA-CNRS, \\ Ecole Polytechnique, F-91128 Palaiseau, France
}
\author{M.E. Povarnitsyn}
\email{povar@ihed.ras.ru}
\affiliation
{
Joint Institute for High Temperatures, RAS, 13 Bld.~2 Izhorskaya str., 125412 Moscow, Russia
}
\author{K.N. Mishchik}
\email{kmishchik@amplitude-systemes.com}
\affiliation
{
Amplitude Syst\`emes, 11 avenue de Canteranne, Cit\'e de la Photonique, F-33600 Pessac, France
}
\author{A. Tanguy}
\email{anne.tanguy@insa-lyon.fr}
\affiliation
{
Universit\'{e} de Lyon, LaMCoS, INSA-Lyon, CNRS UMR5259, F-69621 Villeurbanne Cedex, France
}
\date{\today}  

\begin{abstract}
We report an experimental and numerical study of femtosecond multi-pulse laser-induced densification in vitreous silica ($v$-SiO$_2$) and its signature in Raman spectra. We compare the experimental findings to recently developed molecular dynamics (MD) approach accounting for bond-breaking due to laser irradiation, together with a dynamical matrix approach and bond polarizability model based on first-principle calculations for the estimation of Raman spectra. We observe two stages of the laser-induced densification and Raman spectrum evolution: growth during several hundreds of pulses followed by further saturation. At the medium-range, the network connectivity change in $v$-SiO$_2$ is expressed in reduction of the major ring fractions leading to more compacted structure. With the help of Sen \& Thorpe model, we also study the short-range order transformation and derive the interbonding \hbox{Si--O--Si} angle change from the Raman measurements. Experimental findings are in excellent agreement with our MD simulations, and, hence, support bond-breaking mechanism of laser-induced densification. Thus, our modeling explains well the laser-induced changes both in the short-range order caused by the appearance of Si-coordination defects and medium-range order connected to evolution of the ring distribution. Finally, our findings disclose similarities between sheared-, permanently-densified- and laser-induced-glass and suggest interesting future experiment in order to clarify the impact of the thermo-mechanical history on glasses under shear, cold- and hot-compression, and laser-induced densification.
\end{abstract}
%
%
\maketitle
\section{Introduction}
Comprehension of the processes of laser-induced modification in vitreous silica ($v$-SiO$_2$) and their control remains an important research issue. This concerns the accurate design of the optical properties via local laser-induced refractive index changes (RIC), serving in the fabrication of embedded optical components in fibers and bulk materials~\cite{gattass2008,ams2009,malinauskas2016}. The densification of $v$-SiO$_2$ due to laser irradiation seems reasonable to cause a uniform RIC~\cite{bellouard2008,shcheblanov2016}. The supporting evidence of this mechanism was provided by micro-Raman spectroscopy for both single-~\cite{zoubir2006,shimotsuma2011} and multi-pulse~\cite{Chan2001,streltsov2008,mishchik2013,stoian2013,bellouard2016,stoian2016,Krol2017} experiments.

In these experiments, the analyzes were mainly focused on the strongest band ($\sim437$ cm$^{-1}$) and defects lines D$_{1}$ and D$_{2}$ ($\sim495$ cm$^{-1}$ and~$\sim606$ cm$^{-1}$), usually related to four-membered and three-membered rings (or 4- and 3-fold), respectively. Analysis in terms of 3- and 4-fold rings is still a topic of discussion~\cite{galeener93,alfprl98,huang2013}. It was shown an increase of the intensity of the D$_2$ band, together with a stagnation of the D$_1$ band (in opposition with cold compressed glasses). In Ref.~\cite{shcheblanov2016}, we treated the hypothesis of the bond-breaking mechanism leading to $v$-SiO$_2$ densification under laser irradiation. We addressed to the medium-range order and explained well the behavior of defect line D$_2$ relying on the connection between population of the 3-fold rings and D$_2$ lines. However, all these previous Raman measurements indicating material sensitivity to laser irradiation were not compared to numerical Raman results~\cite{streltsov2008,bressel2011,stoian2013,Krol2017} and no Raman data are available for successive laser-induced experiments in the high-frequency range ($\sim900$--1300 cm$^{-1}$).

In parallel, intensive studies of $v$-SiO$_2$ glass are continuing aimed at studying the role of thermo-mechanical history in densification~\cite{deschamps2014,martinet2015,cornet2017} and plastic-shear~\cite{shcheblanov2015} where, in particular, it is observed a specific sensitivity of Raman spectra in the high-frequency range. In order to infer structural information from Raman spectra in silica glasses, the Sen and Thorpe (ST) analysis~\cite{thorpe1977} relating the position of the characteristic bands to interbonding angle is usually applied~\cite{hehlen2010,weigel2016}. 

Here, we present a systematic analysis of the effect of multi-pulse femtosecond laser on vibrational properties, Raman spectra, medium-range and short-range structure, with the help of numerical calculations compared to experimental measurements of Raman spectra.
\begin{figure} [!ht]
\begin{center}
  \includegraphics[width=8.5cm]{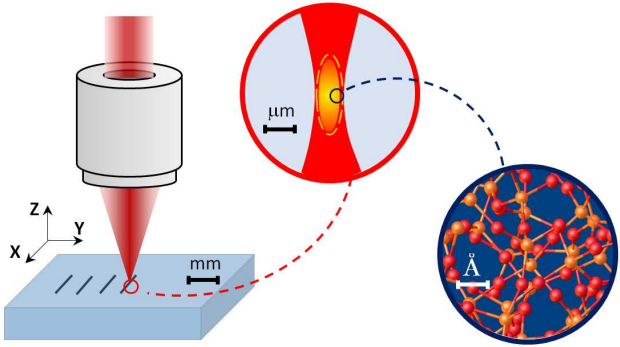}
    \caption{Schematic representation of the femtosecond laser writing. At macro-scale, the laser beam is focused inside the bulk of $v$-SiO$_2$ via objective. At micro-scale, photoionization conditions are realized in the focal area. At nano-scale, the structure of $v$-SiO$_2$ is composed of network-forming species Si (orange) and O (red) undergoing bond-breaking due to photoionization.}
  \label{sketch}
\end{center}
\end{figure}
\section{Methods}
\subsection{Experiment}
The sketch in Fig.~\ref{sketch} shows a multiscale nature of the writing processes involved in the consideration. Femtosecond laser pulses are produced by a regeneratively amplified Ti:sapphire laser system at 800~nm with a nominal pulse duration $\tau_p=130$~fs (FWHM) and an energy 1~$\mu$J. The long working distance $20\times$~microscope objective (Mitutoyo MPlan, NA = 0.42, $f$ = 10~mm) is employed to focus the ultrashort laser pulses into the bulk of silica~\cite{mishchik2013}. A laser beam diameter is less than the objective pupil ($d$ = 5 mm at the level 1/e$^2$), therefore the nominal value of NA is corrected to $\mathrm{NA_{eff}}=0.3$. Polished high-purity synthetic fused silica (Corning 7980-5F, 800-1000 ppm concentration of OH impurities) samples are mounted on a XYZ motion stage. Translation parallel to the laser propagation axis allows to write long waveguiding structures produced with controlled number of pulses. The modifications are produced at 10~kHz laser repetition rate, when the material is thermally relaxed before the new pulse arrival. This is in agreement with other reports indicating that heat accumulation effect in silica plays an important role only at MHz regime of irradiation~\cite{eaton2008}.

Laser modification results in a uniform positive refractive index changes of exposed volume (type I structures~\cite{sugioka2014}), that is confirmed by the waveguiding properties of the written structures and additionally verified using phase contrast microscopy. The Raman spectra of the irradiated samples are recorded with a Horiba Jobin Yvon confocal micro-spectrometer in a backscattering configuration. The laser excitation is performed using a HeCd source at 442~nm wavelength. The arrangement allows for a spatial resolution better than 1~$\mu$m and a spectral resolution of 3~cm$^{-1}$.
\begin{figure} [!ht]
\begin{center}
  \includegraphics[width=8.5cm]{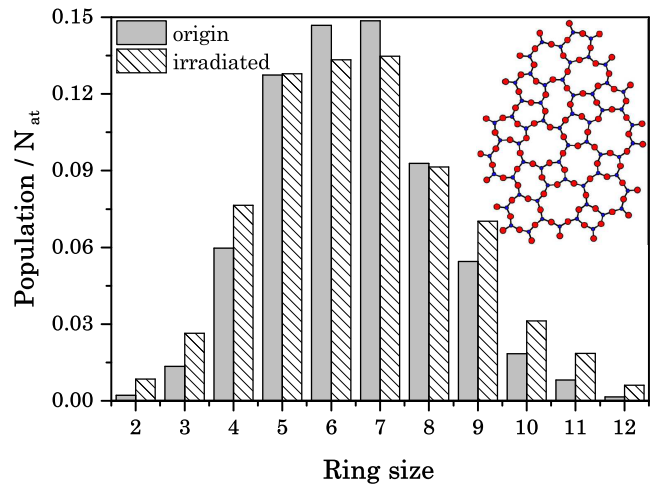}
    \caption{Evolution in the distribution of rings in $v$-SiO$_2$. Grey shaded distribution corresponds to origin samples. Patterned distribution corresponds to samples irradiated by 800 pulses. The results of modeling are obtained by averaging over 10 samples. Number of atoms: $\mathrm{N_{at}} = 8232$. Inset: Two-dimensional schematic diagram of the $v$-SiO$_2$ network with added colors~\cite{zachariasen1932}.}
  \label{rings}
\end{center}
\end{figure}
\subsection{MD simulation}
\label{part:MD}
In our study, the $v$-SiO$_2$ model is prepared following the work~\cite{boris2012}. A system of 8232 atoms (size $\sim$5~nm) is obtained within molecular dynamics (MD) simulation (via \textit{LAMMPS}~\cite{lammps}) using a melt-quench procedure~\cite{binder2011}, and the quenching rate is 5.2$\times10^{13}$~K$\cdot$s$^{-1}$. The equilibration of the liquid, quench and relaxation of the glass are performed classically using the BKS potential~\cite{bks1990} modified by Carr\'{e} \emph{et al.}~\cite{carre2007} (see Appendix~\ref{sec:Etot}). By evolving the model during $10, 20 \ldots 100$~ns at liquid stage, we obtain 10 $v$-SiO$_2$ samples at density of 2.2~g$\cdot$cm$^{-3}$.

In order to consider the interaction of sub-picosecond laser pulses with $v$-SiO$_2$ we introduced a bond-breaking mechanism into the MD scheme~\cite{shcheblanov2016}. The whole simulation cycle of a laser pulse interaction with an MD glass model is as follows. The simulation starts from an instantaneous generation of broken bonds emulating the laser pulse excitation stage. The ionization degree is 0.018\%, corresponding to two broken bonds per sample (or free electron density $1.6\times10^{19}$~cm$^{-3}$). At the second step, the excited electrons transfer energy to the atomic system heating the sample. The temperature saturates with time reaching the maximum value (500~K). Therefore, we evolve our glass sample during the next 5~ps at 500~K. Subsequently, at the third step, a cooling to the room temperature, 300~K, occurs during 100~ps. At the end of the cooling stage, we repair instantly an interaction ability for previously bond-broken Si- and O-atoms. Finally, an annealing at room temperature is applied. We repeat this cycle a required number of times to simulate multi-pulse laser irradiation (up to 1000 pulses).
\begin{figure} [!ht]
\begin{center}
\includegraphics[width=8.2cm]{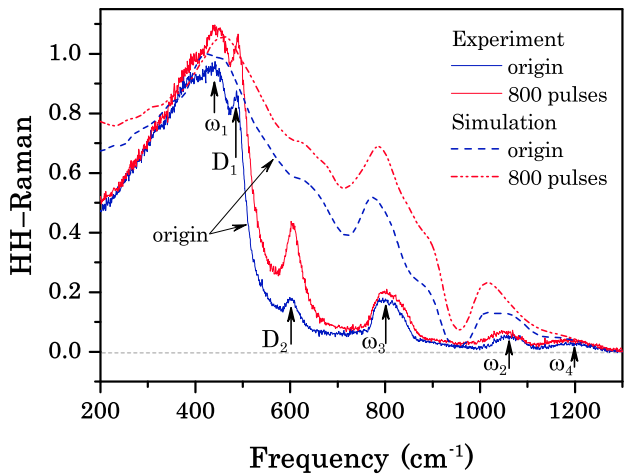}
\caption{HH-Raman spectra in $v$-SiO$_2$. Experimental data: origin sample --- solid blue, after irradiation --- solid red. The simulation results: origin --- dashed blue, after irradiation --- solid dash-dot red. The results of simulation are averaged over 10 configurations. Reference frequencies discussed in the text are $\omega_1=437$~cm$^{-1}$, $\omega_3=800$~cm$^{-1}$, $\omega_2=1060$~cm$^{-1}$, $\omega_4=1200$~cm$^{-1}$.}
  \label{raman}
\end{center}
\end{figure}
\subsection {Raman model}
\label{part:RamanModel}
In our previous study we established that the laser-induced densification of $v$-SiO$_2$ is traced to medium-range changes in topology of the atomic network. These changes consist in an increase of network connectivity caused by the reduction of major ring fractions of six- and seven-membered rings to minor fractions of three- and four-membered rings~\cite{shcheblanov2016}. In Fig.~\ref{rings} we show the evolution in the distribution of rings in $v$-SiO$_2$ upon multi-pulse irradiation. However, it is impossible to directly extract ring statistics or local ring environments in the bulk from experimental measurements. Nonetheless, by applying the Raman spectroscopy enabling to perform the vibrational analysis at $\mu$m-scale we can infer the information at medium-range and short-range order.

Here, we briefly outline the formulation that we use for the calculation of Raman activities. We focus only on first-order processes, which involve a single phonon excitation. In the Stokes process, in which a vibrational excitation is created by an incoming photon, we express the total power cross-section as (in esu units)~\cite{cardona1982}:
\begin{equation}
  \mathcal{I}^{P}(\omega) = \frac{2\pi\hbar}{\omega}\;\frac{g(\omega)(\omega_{L}-\omega)^4}{V^{-1}c^4}\sum_{n}\mathcal{I}_n\,\delta(\omega-\omega_n),
  \label{tram}
\end{equation}
\begin{figure}[!ht]
\begin{center}
  \includegraphics[width=8.5cm]{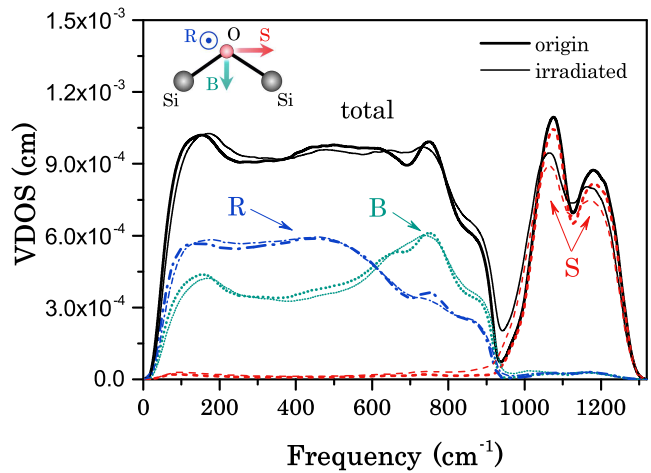}
    \caption{The partial VDOS for the projections onto the vibrations of \hbox{Si--O--Si} structural units: stretching --- dashed red, bending --- dotted green, rocking --- dash-dot blue and total --- solid black. Origin sample --- thick curves, after irradiation by 800 pulses --- thin curves. Inset shows the relative motion of the oxygen atoms decomposed into stretching (S), bending (B), and rocking (R) components.}
  \label{modesVDOSBRS}
\end{center}
\end{figure}
where the index $n$ labeling the vibrational modes runs from 1 to $3N_{at}$, $N_{at}$ is the total number of atoms in the model (8232 atoms), $\omega_L$ is the frequency of the incoming photon, $c$ is the speed of light, $V$ is the volume of the scattering sample, $g(\omega)=n_B(\omega)+1$, and $n_B(\omega)$ is the boson factor. In experimental set-ups, it is customary to record the Raman spectra in the horizontal-horizontal (HH) configuration in which the polarization of the outgoing photons is respectively parallel to the ingoing photon polarization~\cite{bruesch1986}. Using the isotropy of disordered solids, we express the contribution of the $n$-th mode, $\mathcal{I}_n$ to the HH-Raman spectra as~\cite{cardona1982}:
\begin{equation}
  \mathcal{I}_{n}^{HH} = a^{2}_{n} + \frac{4}{45}b^{2}_{n},
  \label {pram}
\end{equation}
where $a_n$ and $b_n$ are obtained from:
%
\begin{gather}
  a_{n} = \frac {1}{3}\sum_i^3\mathcal{R}^{n}_{ii},\\
%
  b^{2}_{n} = \sum_{i<j}^3\left\{\frac{1}{2}\left(\mathcal{R}^{n}_{ii}-\mathcal{R}^{n}_{jj}\right)^2+3\left(\mathcal{R}^{n}_{ij}\right)^2\right\}.
  \label{anbn}
\end{gather}
%
The Raman susceptibility tensors $\mathcal{R}^n_{ij}$ are given by~\cite{cardona1982}:
\begin{equation}
  \mathcal{R}^n_{ij} = {\sqrt{V}}\sum_{I,k} {\frac{\partial \chi_{ij}}{\partial R_{I k}}\frac{\xi^{n}_{I k}}{\sqrt{M_I}}},
  \label{ramten}
\end{equation}
where $\mathbold{\chi}$ is the electric polarizability tensor, the capital Latin indices run over the atoms, the lowercase Latin indices are the three Cartesian directions, ${\bf R}_I=(R_{I1},R_{I2},R_{I3})$ and $M_I$ are the position and the atomic mass of atom $I$, respectively. In order to compute $\mathbold{\chi}$, we apply the bond polarizability model~\cite{wolk1941,wolk1944} (see Appendix~\ref{sec:bpm}). For a model system, the vibrational frequencies $\omega^2_n$ and their associated eigenmodes $\mathbold{\xi}^{n}_I$ are found by solving the set of linear equations:
\begin{equation}
  \sum_{J j} D_{I i,J j} \xi^{n}_{J j} = \omega^{2}_{n}\xi^{n}_{I i},
  \label{eigen}
\end{equation}
where $\bf {D}$ is the dynamical matrix, which is defined by:
%
\begin{equation}
  D_{I i,J j} = \frac{1}{\sqrt{M_I M_J}} \frac{\partial^2E_{tot}}{\partial R_{I i}\partial R_{J j}},\,\,\mathrm{for}\,\,I\neq J 
  \label{dynmatne}
\end{equation}
\begin{equation}
  D_{I i,I j} = - \sum_{J\neq I} \frac{1}{M_I} \frac{\partial^2E_{tot}}{\partial R_{I i}\partial R_{J j}},
  \label{dynmat}
\end{equation}
where $E_{tot}$ is the global potential energy of the system (see Appendix~\ref{sec:Etot}).

The result of the Raman simulations of $v$-SiO$_2$ are presented in Fig.~\ref{raman}. One can see that the main characteristics of the experimental spectra are recovered within our semi-classical approximation. In the Raman spectrum, we can recognize well the main band ($\sim$~400--550 cm$^{-1}$) as well as the high-frequency bands, in particular, those located at $\sim$~800 cm$^{-1}$, $\sim$~1060 cm$^{-1}$, and $\sim$~1200 cm$^{-1}$, that are well reproduced by our simulation.

The assignment of the vibration modes of $v$-SiO$_2$ is well documented~\cite{taraskin97,taraskin2016,alf97,umari03}. Performing projectional analysis (see Appendix~\ref{sec:vdos}), we decompose the vibrational density of states (VDOS) for relative motions of the oxygen atoms into stretching, bending, and rocking components (see Fig.~\ref{modesVDOSBRS}). The VDOS reveals pure stretching nature of the high-frequency doublet zone ($\sim$~950--1300~cm$^{-1}$), the bending and rocking modes are in the range \hbox{$\sim$~0--900~cm$^{-1}$}. In all the spectra~Figs.~\ref{raman} and~\ref{modesVDOSBRS}, the high-frequency part of the spectra ($\gtrsim$~950~cm$^{-1}$) almost exclusively results from stretching vibrations. As far as the HH-Raman spectrum is concerned, Umari and Pasquarello showed that the bending motions dominate the rest of the spectrum ($\sim$~100--900~cm$^{-1}$), whereas the contribution of rocking vibrations is suppressed with respect to their weight in the VDOS~\cite{umari03}.
\section{Results and discussion}
\begin{figure} [!ht]
\begin{center}
  \includegraphics[width=8.0cm]{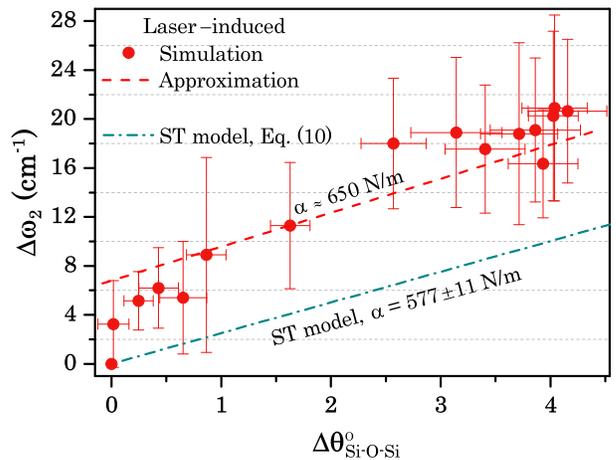}
    \caption{Simulation of laser-induced HH-Raman $\omega_2$-shift versus mean \hbox{Si--O--Si} angle change --- filled red circles (half width at half maximum fitting). The dashed red line shows the approximation of the simulation data. The results of modeling are averaged over 10 samples. The dash-dot green line shows $\omega_2$-shift based on the ST model~\cite{thorpe1977}, Eq.~(\ref{dw2dtheta}), at parameters: central-force constant $\alpha = 577$~N/m (calculated using measured Raman characteristic frequency $\omega_2$, see Fig.~\ref{raman}) and mean \hbox{Si--O--Si} angle, $\vartheta=147.7\pm3.8^\circ$~\cite{trease2017}.}
  \label{shift_angle}
\end{center}
\end{figure}
Comparing the experimental and modeling Raman spectra from Fig.~\ref{raman}, we highlight that the laser-induced changes in all bands have the same behavior. In particular, one can observe a raise and a slight shift to the right of the main band $\omega_1$; the $\omega_3$-band has a weaker raise and a slight shift to the right. On the contrary, we can indicate the more significant changes in the high-frequency doublet zone, a shift to the left of $\omega_2$ and $\omega_4$ bands. In the experimental spectrum, one can see a significant gain of defect lines D$_1$ and D$_2$, whereas our simulation suffers due to the lack of resolution in these areas. However, the D$_1$ and D$_2$ Raman defect lines are usually associated with 4- and 3-fold rings in the structure of $v$-SiO$_2$~\cite{alfprl98}. Since these lines provide direct information on the concentration of these rings~\cite{umari2003}, we characterize the medium-range structure (network connectivity) by analyzing the ring distribution instead of a direct Raman analysis of the defect lines. Our previous work was dedicated to this point~\cite{shcheblanov2016}. It is also interesting to note, that the global increase in the Raman intensity in the intermediate frequency range, and the shift of the high-frequency bands in both Raman and VDOS (see Fig.~\ref{modesVDOSBRS}) spectra share more similarities with the Raman and VDOS signatures of plastic-shear~\cite{shcheblanov2015} and permanently densified glass~\cite{hiramatsu1996,hehlen2010,weigel2016}.

In order to explain the changes in the Raman spectra upon multi-pulse laser irradiation, we exploit a simple central-force ST model to describe the dynamics of covalently bonded networks~\cite{thorpe1977}. Ref.~\cite{thorpe1977} assumed that the vibrations of the silica network can be described using \hbox{Si--O--Si} units with only one force constant $\alpha$ defined by the \hbox{Si--O} bond. The following equations were proposed to describe the positions of the vibrational bands as a function of interbonding angle~$\vartheta$:
\begin{subequations}
\begin{align}
\omega^{2}_{1}=\frac{\alpha}{M_{\mathrm{O}}} \left(1+\mathrm{cos}\,\vartheta \right), \label{om1} \\
%
\omega^{2}_{2}=\frac{\alpha}{M_{\mathrm{O}}} \left(1-\mathrm{cos}\,\vartheta \right), \label{om2} \\
%
\omega^{2}_{3}=\omega^{2}_{1}+\frac{4\alpha}{3M_{\mathrm{Si}}}, \label{om3} \\
%
\omega^{2}_{4}=\omega^{2}_{2}+\frac{4\alpha}{3M_{\mathrm{Si}}},  \label{om4}
\end{align}
\end{subequations}
where $M_{\mathrm{O}}$ and $M_{\mathrm{Si}}$ are the masses of oxygen and silicon atoms, respectively.
\begin{figure*} [!ht]
\begin{center}
  \includegraphics[width=8.5cm]{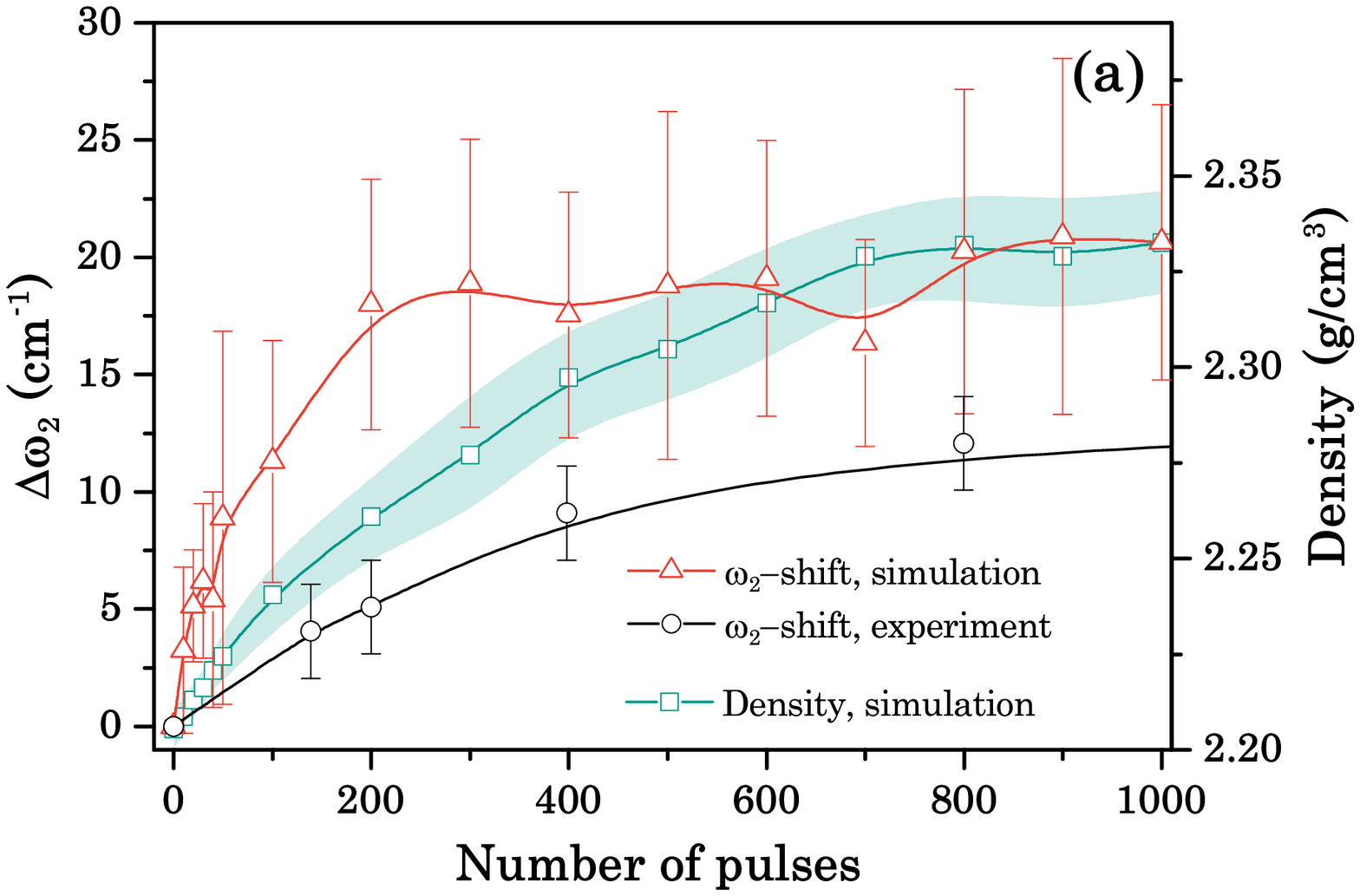}\qquad\qquad
  \includegraphics[width=7.2cm]{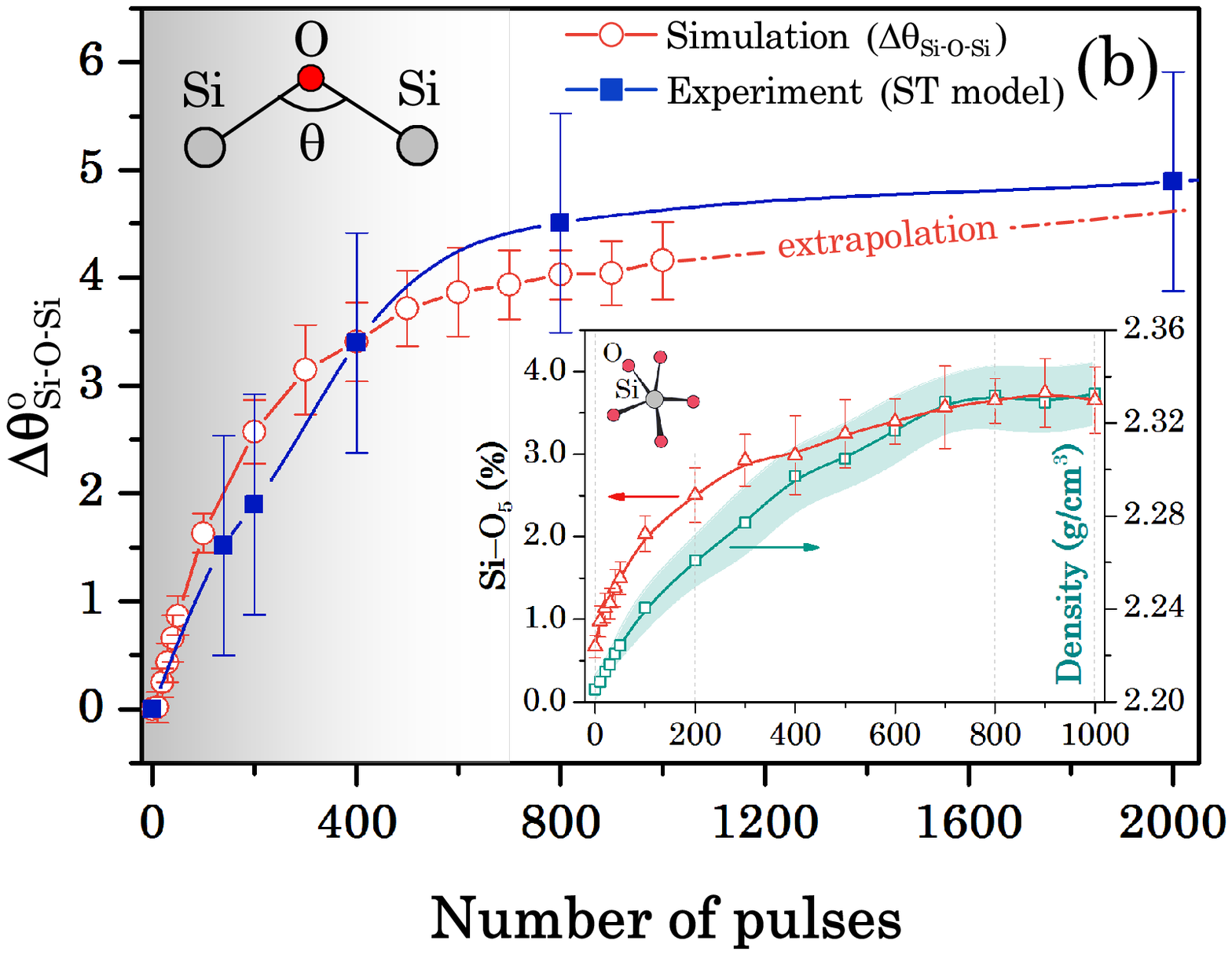}
    \caption{(a) Absolute value $\mathrm{\Delta}\omega_2$ of the laser-induced HH-Raman $\omega_2$-shift as a function of pulse number: experiment --- empty black circles; simulation --- empty red triangles, compared to density change obtained in simulation --- empty green squares with density dispersion depicted by the shaded green area. (b) \hbox{Si--O--Si} angle change versus pulse number. Experimental results via ST model --- filled blue squares; results of simulation --- empty red circles. Inset: Change of number of 5-coordinated Si-atoms at degree of ionization 0.018\% in simulation as a function of pulse number --- empty red triangles. Corresponding density --- empty green squares with error bar depicted by shaded light green area. The results of modeling are averaged over 10 samples.}
  \label{delta_angle_pulses}
\end{center}
\end{figure*}

To extract information about the angle change from Raman spectra applying the ST model, one needs to determine the central-force constant. By using the measured Raman frequencies $\omega_{i}$ (see Fig.~\ref{raman}) and mean \hbox{Si--O--Si} angle, $\vartheta=147.7^\circ$, taken from~\cite{malfait2008,trease2017}, the force constants $\alpha_{1-4}$ are calculated by Eqs.~(\ref{om1})--(\ref{om4}), respectively. The values of $\alpha$ are presented in Table~\ref{Tab1}. It can be seen that $\alpha_1(\omega_1)\approx1170$~~N/m differs significantly from other values. This is probably due to a wide width of the asymmetric $\omega_1$-band ($\sim$100--500~cm$^{-1}$) and a broad \hbox{Si--O--Si} angle distribution ($\sim$135--160$^\circ$~\cite{trease2017}) that leads to an ambiguity in the calculation of $\alpha_1$. However, if we take instead of the frequency $\omega_1=437$~cm$^{-1}$ (corresponding to the maximum of the band), the frequency $\omega_{g}\approx340$~cm$^{-1}$ corresponding to the center of gravity of the $\omega_1$-band, then $\alpha_{1}(\omega_g)\approx710$~N/m. Thus, all $\alpha_{1-4}$ according to the ST model are in quite good agreement with each other.
\begin{table} [!bt]
\setlength{\tabcolsep}{4.1pt}
\begin{tabular}{c c c c c c c}
\toprule
\\
Force const. & $\alpha_{1} (\omega_1)$ & $\alpha_{1}(\omega_g)$ & $\alpha_{2}$ & $\alpha_{3}$ & $\alpha_{4}$ & $\bar \alpha$ \\
\\
\hline \\
N/m & 1170.1 & 710 & 577.3 & 558.8 & 394.0 & 560.0 \\
\\
\toprule
\end{tabular}
\caption{Force constants. $\bar \alpha=(\alpha_{1}(\omega_g)+\alpha_{2}+\alpha_{3}+\alpha_{4})/4$, where $\alpha_{1-4}$ are calculated by Eqs.~(\ref{om1})--(\ref{om4}), respectively.}
\label{Tab1}
\end{table}

As shown in~\cite{geissberger83}, the interbonding angle change $\mathrm{\Delta}\vartheta$ can be related to the $\omega_2$-shift by differentiating Eq.~(\ref{om2}) to obtain:
\begin{equation}
  \mathrm{\Delta}\omega_{2}=\frac{\alpha}{2M_{\mathrm{O}}}\frac{\mathrm{sin}\,\vartheta\mathrm{\Delta}\vartheta}{\omega_{2}}.
  \label{dw2dtheta}
\end{equation}
Our simulation and experimental measurements provide good resolution of the Raman $\omega_2$-band. Furthermore, we choose the $\omega_2$-band to treat the experimental measurements because the corresponding $\alpha_2$ is closer to the mean value of central-force constant $\bar\alpha$, see Table~\ref{Tab1}.

In Fig.~\ref{shift_angle} we present the simulation results of $\omega_2$-shift as a function of mean interbonding angle change obtained by irradiation with different number of pulses. In result, we can see that our simulations predict the linear behavior confirming the ST model, see Eq.~(\ref{dw2dtheta}). Moreover, the central-force constant values obtained from the simulation ($\alpha\approx650$~N/m) and calculated using the measured Raman characteristic frequency $\omega_2$ ($\alpha\approx577$~N/m) are in good agreement. Thus, we demonstrate that the ST model is applicable to describe the Raman dynamics upon multiple laser irradiation.

The dynamics of $\mathrm{\Delta}\omega_2$ with the pulse number is shown in Fig.~\ref{delta_angle_pulses}a. On can reveal two stages of dynamics, the growth during several hundreds of pulses followed by saturation, both in simulation and experiment. Despite the averaging over 10 samples, we can mention the sharp rise of the $\omega_2$-shift in the very beginning and noticeable dispersion in simulation that we ascribe to the size of the models (see Subsection~\ref{part:MD}). This size effect will be reduced in our further investigations by increasing the number of atoms in the samples. We can also observe two stages, growth and saturation, in the densification dynamics, that correlates quite well with the $\omega_2$-shift, however, it is rather smooth compared to the $\omega_2$-shift.

Applying the ST model to the $\omega_2$-shift experimental findings from Fig.~\ref{delta_angle_pulses}a, we extract the corresponding angle change. One can see that the interbonding angle change follows the $\omega_2$-shift and also reproduces the two-stage dynamics: the growth during $\sim600$ pulses followed by a saturation, see Fig.~\ref{delta_angle_pulses}b. The comparison shows an excellent agreement between experiment and MD simulation. Since we reveal the densification, we guess that laser-induced rearrangements should lead to transformations of $v$-SiO$_2$ network both in the medium-range~\cite{shcheblanov2016} and short-range order. The short-range structure is usually characterized by coordination number~\cite{trave2002,zeidler2014}. We found here, a correlation between the growth of five-coordinated Si-atoms and $v$-SiO$_2$ densification upon laser irradiation, see inset in Fig.~\ref{delta_angle_pulses}b. Similar behavior was observed in permanently densified silica glass at high pressure~\cite{trave2002,benmore2010,zeidler2014}. Thus, we relate the \hbox{Si--O--Si} angle decline to the increase of network connectivity caused by the reduction of the major ring fractions~\cite{shcheblanov2016} and the increase in coordination defects (five-coordinated Si-atoms) due to multi-pulse laser irradiation.
\section{Conclusions}
In summary, our systematic Raman calculation shows that upon multi-pulse laser irradiation the SiO$_2$ glass undergoes several successive transformations both in short-range and medium-range order. The changes in the Raman measurements are well described by our simulations within the numerical accuracy. This reinforces our previously developed model of bond-breaking for laser irradiation~\cite{shcheblanov2016}. In experiment and simulation, two stages of the laser-induced densification and Raman spectrum evolution are observed: growth during several hundreds of pulses followed by further saturation. At the medium-range, the network connectivity is expressed in reduction of the major ring fractions leading to more compacted structure. By using the Raman measurements and the ST model, we highlight the short-range transformation by extracting the dynamics of \hbox{Si--O--Si} angle change. These results are in excellent agreement with our simulation results. In addition, we show a correlation between the growth of five-coordinated Si-atoms and densification due to laser irradiation. Thus, we conclude that the laser-induced densification of $v$-SiO$_2$ is related to the changes in the short-range order caused by the appearance of Si-coordination defects and medium-range order connected to evolution of the ring distribution. These findings disclose similarities between laser-induced- and permanently-densified glass~\cite{deschamps2014,cornet2017}. Moreover, our preliminary analysis shows more generally that the sensitivity of the Raman spectra to pressure variation depends strongly on the samples preparation protocol. We also note that the global increase in the Raman intensity in the intermediate frequency range, and the shift of the high-frequency bands share more similarities with the Raman signature of plastic-shear~\cite{shcheblanov2015}. Therefore, our simulation results suggest interesting future experiment in order to clarify the impact of the thermo-mechanical history on glasses under shear, cold- and hot-compression, and laser-induced densification.
\acknowledgments
M.E. Povarnitsyn was supported by the Russian Science Foundation, grant No. 16-19-10700. The authors thank LaHC (Saint-Etienne) and ILM (Lyon) for stimulating discussions, and Intel team (G. Fedorov, V. Chumakov and I. Sokolova) for support with the FEAST Eigenvalue Solver. N.S. Shcheblanov and A. Tanguy acknowledge the French Research National Agency program ANR MECASIL (ANR-12-BS04-0004) and BQR financing from LaMCoS.
\appendix
\addcontentsline{toc}{section}{Appendix}
\section{BKSW potential}
\label{sec:Etot}
The equilibration of the liquid, quench and relaxation of the glass are performed classically using the BKSW potential~\cite{carre2007}, which is a modified version of the van Beest, Kramer and van Santen (BKS) potential~\cite{bks1990}. It can be described as a two-body potential:
\begin{align}
  \Phi^{BKSW}_{\alpha \beta}(R) = \Phi^{Coul}_{\alpha  \beta}(R) + \Phi^{Buck}_{\alpha  \beta}(R),
  \label{bksw}
\end{align}
where $\alpha$ and $\beta$ are the types of atoms (O or Si), and $R$ is the distance between them.
\begin{align}
  \Phi^{Coul}_{\alpha  \beta}(R) = q_{\alpha}q_{\beta}e^{2}V_{W}(R)G_{W}(R),
  \label{coul}
\end{align}
with
\begin{gather}
  V_{W}(R) = \left(\frac{1}{R} - \frac{1}{R_{c,W}} \right) + \frac{1}{R^{2}_{c,W}} (R - R_{c,W}), \\
  G_{W}(R) = \exp \left(-\frac{\gamma^{2}_{W}}{(R - R_{c,W})^{2}} \right).
  \label{wolfeq}
\end{gather}
\begin{align}
  \Phi^{Buck}_{\alpha  \beta}(R)& = \left[ A_{\alpha  \beta} \left( e^{-\frac{R}{\rho_{\alpha \beta}}} - e^{-\frac{R_{c,sh}}{\rho_{\alpha \beta}}} \right) \right.  \nonumber \\
 & \qquad  \left. - C_{\alpha \beta} \left( \frac{1}{R^{6}} - \frac{1}{R^{6}_{c,sh}} \right) \right] G_{sh}(R),
  \label{buck}
\end{align}
with
\begin{equation}
  G_{sh}(R) = \exp \left(-\frac{\gamma^{2}_{sh}}{(R - R_{c,sh})^{2}} \right),
  \label{tranc}
\end{equation}
where $\gamma_{sh} = \gamma_{W} = 0.5$ \AA, $R_{c,W} = 10.17$ \AA, and $R_{c,sh} = 5.5$ \AA.
We also add a strong and regular repulsive part at short range $(R < R_{inf})$ to avoid the collapse of atoms at high pressure, or high temperatures. The added repulsive part has the following form:
\begin{equation}
  \Phi^{Rep}_{\alpha  \beta}(R) = \left ( \frac{D_{\alpha \beta}}{R} \right)^{12}+E_{\alpha \beta}R+F_{\alpha \beta}.
  \label{rep}
\end{equation}
$D_{\alpha \beta}$, $E_{\alpha \beta}$, and $F_{\alpha \beta}$ have been adjusted in order to have the continuity of the potential and its first, and second derivatives. The parameters of this potential are tabulated in Table~\ref{Tab3}. The total energy, which is used to compute the Dynamical Matrix, can be written:
\begin{equation}
  E_{tot}=\sum_{I<J} \Phi^{BKSW}_{\alpha_I \beta_J}(R)+\Phi^{Rep}_{\alpha_I \beta_J}(R),
\end{equation}
and here $R$ is the distance between atoms $I$ and $J$.
\begin{table} [!ht]
\resizebox{8.5cm}{!} {
\begin{tabular}{c c c c c}
\toprule
\\
$   $ & $A_{\alpha \beta}$ (eV) & $\rho_{\alpha \beta}$ (\AA) & $C_{\alpha \beta}$ (eV$\cdot$\AA$^{6}$) & $D_{\alpha \beta}$ (\AA$\cdot$eV$^{-12}$)  \\
\\
\hline
\\
O-O & 1388.773 & 0.3623 & 175.0 & 1.51166281 \\
Si-O & 18003.7572 & 0.2052 & 133.5381 & 1.42402882 \\
Si-Si & 872360308.1 & 0.0657 &23.299907 & 0.0 \\
\\
\hline
\\
 $   $& $E_{\alpha \beta}$ (eV$\cdot$\AA$^{-1}$) & $F_{\alpha \beta}$ (eV) & $R_{inf}$ (\AA) & $ $  \\
 \\
\hline
\\
O-O & -14.97811134 & 39.0602602165 & 1.75 & \\
Si-O & -3.24749265 & -15.86902056 & 1.27 & \\
Si-Si & 0.0 & 0.0 & 0.0 & \\
\\
\toprule
\end{tabular}
}
\caption{Parameters of the empirical potential used to model the silica glass.}
  \label{Tab3}
\end{table}
\section{Bond polarizability model}
\label{sec:bpm}
The bond polarizability model (BPM)~\cite{wolk1941,wolk1944} has successfully been applied for the calculation of Raman intensities in a large variety of systems \cite{cardona1982}. In this approach, the polarizability is modeled in terms of bond contributions:
\begin{equation} \label {ex10}
  \chi_{ij}(I) = \frac{1}{V}\sum_J\alpha_{ij}(I,J),
\end{equation}
where the polarizability tensors $\alpha_{ij}(I,J)$:
\begin{equation}
  \alpha_{ij}=\frac{1}{3}(2\alpha_p+\alpha_l)\delta_{ij} + (\alpha_l-\alpha_p)\left(\frac{R_iR_j}{\vert {\bf {R}}\vert^2}-\frac{1}{3}\delta_{ij}\right),
\end{equation}
where ${\bf{R}} = {\bf{R}}_I-{\bf {R}}_J$ is a vector which defines the direction and the distance of a pair of nearest neighbor atoms at sites ${\bf{R}}_I$ and ${\bf{R}}_J$. The parameters $\alpha_l$ and $\alpha_p$ correspond to the longitudinal and perpendicular bond polarizability, respectively.

The BPM further assumes that the bond polarizabilities $\alpha_l$ and $\alpha_p$ only depend on the length of the bond. Thus the derivative of the local bond polarizability with respect to the relative displacement of the atoms $I$ and $J$ yields:
\begin{multline} \label {ex11}
\frac{\partial \alpha_{ij}(I,J)}{\partial R_{I k}} = \frac{1}{3}\left(2\alpha'_{p}+\alpha'_{l}\right)\delta_{ij}\hat{R}_k \\
+\left(\alpha'_l-\alpha'_p\right)\left(\hat{R}_i\hat{R}_j-\frac{1}{3}\delta_{ij}\right)\hat{R}_k \\
+\frac{(\alpha_l-\alpha_p)}{R}\left(\delta_{ik}\hat{R}_j+\delta_{jk}\hat{R}_i-2\hat{R}_i\hat{R}_j\hat{R}_k\right),
\end{multline}
where $\bf{\hat{R}}$ is a unit vector along $\bf{R}$, $\alpha'_l$ and $\alpha'_p$ are the derivatives of the bond polarizabilities with respect to the bond length ($\alpha'_{l,p} = (\partial \alpha_{l,p} / \partial R ) |_{R=R_{0}} $ and $R_{0}$ is a typical distance). Therefore, when one type of bond occurs, the BPM is completely defined by three parameters: $2\alpha'_p + \alpha'_l$, $\alpha'_l - \alpha'_p$, and $(\alpha_l - \alpha_p) / R$. We use the parameters of the BPM already derived in Refs.~\cite{alf2001,giacomazzi2009}, whose values are summarized in Table~\ref{Tab2}.
\begin{table} [!ht]
\setlength{\tabcolsep}{8pt}
\begin{tabular}{ c c c c }
\toprule
\\
Parameter & $2\alpha'_{p} + \alpha'_l$ & $\alpha'_{l} - \alpha'_{p}$ & $(\alpha_{l} - \alpha_{p}) / R$  \\
\\
\hline \\
$(4\pi)^{-1}\cdot$Bohr$^{-1}$ & 0.771 & 0.196 & 0.056 \\
\\
\toprule
\end{tabular}
\caption{Bond polarizability model parameters.}
\label{Tab2}
\end{table}
\section{Vibrational density of states}
\label{sec:vdos}
The frequencies $\omega^2_n$ and the corresponding normalized eigenmodes $\mathbold{\xi}^n_I$ are obtained by diagonalizing the dynamical matrix. The FEAST solver integrated into Intel MKL is used for the diagonalization~\cite{intel_mkl}. The associated atomic displacements are given by:
\begin{equation}
  {\bf u}^n_I = \frac{\mathbold{\xi}^n_I}{\sqrt{M_I}}.
  \label {dis}
\end{equation}
The index $n$ labeling the vibrational modes runs from 1 to $3N_{at}$, $N_{at}$ is the total number of atoms in the model.

The structure of $v$-SiO$_2$ consists of corner-shared tetrahedral SiO$_4$ units. These units are connected to each other via bridging oxygen atoms. Since oxygen vibrations give the prominent contribution to the Raman spectra, we further decompose this contribution according to three orthogonal directions which characterize the local environment of each oxygen atom~\cite{bell72}. Considering the plane containing the silicon atoms to which a given oxygen atom is bonded, we defined the three directions as in~\cite{bell72,alfprb98}. We took the first direction orthogonal to the \hbox{Si--O--Si} plane (rocking), the second one along the bisector of the \hbox{Si--O--Si} angle (bending), and the third one orthogonal to the two previous ones (stretching), see the sketch in Fig.~\ref{modesVDOSBRS}. The decomposition is carried out by projecting the displacements ${\bf{u}}^{n}_{I}$ onto these directions prior to the calculation of the VDOS spectra, i.e. ${\bf{u}}^{n}_{I}={\bf{u}}^{n}_{Ir}+{\bf{u}}^{n}_{Ib}+{\bf{u}}^{n}_{Is}$ and related to rocking (${\bf{u}}^{n}_{Ir}$), bending (${\bf{u}}^{n}_{Ib}$), and stretching (${\bf{u}}^{n}_{Is}$) motions. Bearing in mind that the two silicon atoms Si$_{1(I)}$ and Si$_{2(I)}$ that are neighbors of the oxygen atom $I$ move as well, only the relative motion of the oxygen atom is decomposed, so that $\widetilde{\bf{u}}^{n}_{I} = {\bf{u}}^{n}_{I}-({\bf{u}}^{n}_{\mathrm{Si}_{1}(I)}+{\bf{u}}^{n}_{\mathrm{Si}_{2}(I)})/2$ is the displacement of oxygen atom $I$ relative to the average displacement of its nearest silicon neighbors. Total VDOS decomposes $Z(\omega)$ into the stretching, bending, and rocking components, $Z(\omega) = Z_r(\omega)+Z_b(\omega)+Z_s(\omega)$:
\begin{equation}
  Z_{r,b,s}(\omega) = \frac{1}{3N_{at}} \sum_n {{\left | \widetilde{r}^{\,n} \right |}^2_{r,b,s} \delta \left(\omega-\omega_n\right) },
\end{equation}
where the squared average displacement can be calculated according to the following expression:
\begin{equation}
  {\left | \widetilde{r}^{\,n} \right |}^2_{r,b,s}=\frac{1}{\sum_I {{\left | \widetilde{\bf{u}}^{n}_I \right |}^2 }} \sum_I^{N_{\mathrm{O}}} {{\left | \widetilde{\bf{u}}^{n}_I \right |}^2_{r,b,s} },
\end{equation}
where $N_\mathrm{O}$ is the number of oxygen atoms.
%

\begin{thebibliography}{54}%
\makeatletter
\providecommand \@ifxundefined [1]{%
 \@ifx{#1\undefined}
}%
\providecommand \@ifnum [1]{%
 \ifnum #1\expandafter \@firstoftwo
 \else \expandafter \@secondoftwo
 \fi
}%
\providecommand \@ifx [1]{%
 \ifx #1\expandafter \@firstoftwo
 \else \expandafter \@secondoftwo
 \fi
}%
\providecommand \natexlab [1]{#1}%
\providecommand \enquote  [1]{``#1''}%
\providecommand \bibnamefont  [1]{#1}%
\providecommand \bibfnamefont [1]{#1}%
\providecommand \citenamefont [1]{#1}%
\providecommand \href@noop [0]{\@secondoftwo}%
\providecommand \href [0]{\begingroup \@sanitize@url \@href}%
\providecommand \@href[1]{\@@startlink{#1}\@@href}%
\providecommand \@@href[1]{\endgroup#1\@@endlink}%
\providecommand \@sanitize@url [0]{\catcode `\\12\catcode `\$12\catcode
  `\&12\catcode `\#12\catcode `\^12\catcode `\_12\catcode `\%12\relax}%
\providecommand \@@startlink[1]{}%
\providecommand \@@endlink[0]{}%
\providecommand \url  [0]{\begingroup\@sanitize@url \@url }%
\providecommand \@url [1]{\endgroup\@href {#1}{\urlprefix }}%
\providecommand \urlprefix  [0]{URL }%
\providecommand \Eprint [0]{\href }%
\providecommand \doibase [0]{http://dx.doi.org/}%
\providecommand \selectlanguage [0]{\@gobble}%
\providecommand \bibinfo  [0]{\@secondoftwo}%
\providecommand \bibfield  [0]{\@secondoftwo}%
\providecommand \translation [1]{[#1]}%
\providecommand \BibitemOpen [0]{}%
\providecommand \bibitemStop [0]{}%
\providecommand \bibitemNoStop [0]{.\EOS\space}%
\providecommand \EOS [0]{\spacefactor3000\relax}%
\providecommand \BibitemShut  [1]{\csname bibitem#1\endcsname}%
\let\auto@bib@innerbib\@empty
\bibitem [{\citenamefont {Gattass}\ and\ \citenamefont
  {Mazur}(2008)}]{gattass2008}%
  \BibitemOpen
  \bibfield  {author} {\bibinfo {author} {\bibfnamefont {R.~R.}\ \bibnamefont
  {Gattass}}\ and\ \bibinfo {author} {\bibfnamefont {E.}~\bibnamefont
  {Mazur}},\ }\bibfield  {title} {\emph {\bibinfo {title} {Femtosecond laser
  micromachining in transparent materials},\ }}\href {\doibase
  10.1038/nphoton.2008.47} {\bibfield  {journal} {\bibinfo  {journal} {Nat.
  Photonics}\ }\textbf {\bibinfo {volume} {2}},\ \bibinfo {pages} {219}
  (\bibinfo {year} {2008})}\BibitemShut {NoStop}%
\bibitem [{\citenamefont {Ams}\ \emph {et~al.}(2009)\citenamefont {Ams},
  \citenamefont {Marshall}, \citenamefont {Dekker}, \citenamefont {Piper},\
  and\ \citenamefont {Withford}}]{ams2009}%
  \BibitemOpen
  \bibfield  {author} {\bibinfo {author} {\bibfnamefont {M.}~\bibnamefont
  {Ams}}, \bibinfo {author} {\bibfnamefont {G.~D.}\ \bibnamefont {Marshall}},
  \bibinfo {author} {\bibfnamefont {P.}~\bibnamefont {Dekker}}, \bibinfo
  {author} {\bibfnamefont {J.~A.}\ \bibnamefont {Piper}}, \ and\ \bibinfo
  {author} {\bibfnamefont {M.~J.}\ \bibnamefont {Withford}},\ }\bibfield
  {title} {\emph {\bibinfo {title} {Ultrafast laser written active devices},\
  }}\href {\doibase 10.1002/lpor.200810050} {\bibfield  {journal} {\bibinfo
  {journal} {Laser Photon. Rev.}\ }\textbf {\bibinfo {volume} {3}},\ \bibinfo
  {pages} {535} (\bibinfo {year} {2009})}\BibitemShut {NoStop}%
\bibitem [{\citenamefont {Malinauskas}\ \emph {et~al.}(2016)\citenamefont
  {Malinauskas}, \citenamefont {{\v{Z}}ukauskas}, \citenamefont {Hasegawa},
  \citenamefont {Hayasaki}, \citenamefont {Mizeikis}, \citenamefont
  {Buividas},\ and\ \citenamefont {Juodkazis}}]{malinauskas2016}%
  \BibitemOpen
  \bibfield  {author} {\bibinfo {author} {\bibfnamefont {M.}~\bibnamefont
  {Malinauskas}}, \bibinfo {author} {\bibfnamefont {A.}~\bibnamefont
  {{\v{Z}}ukauskas}}, \bibinfo {author} {\bibfnamefont {S.}~\bibnamefont
  {Hasegawa}}, \bibinfo {author} {\bibfnamefont {Y.}~\bibnamefont {Hayasaki}},
  \bibinfo {author} {\bibfnamefont {V.}~\bibnamefont {Mizeikis}}, \bibinfo
  {author} {\bibfnamefont {R.}~\bibnamefont {Buividas}}, \ and\ \bibinfo
  {author} {\bibfnamefont {S.}~\bibnamefont {Juodkazis}},\ }\bibfield  {title}
  {\emph {\bibinfo {title} {Ultrafast laser processing of materials: from
  science to industry},\ }}\href {\doibase 10.1038/lsa.2016.133} {\bibfield
  {journal} {\bibinfo  {journal} {Light: Sci. Appl.}\ }\textbf {\bibinfo
  {volume} {5}},\ \bibinfo {pages} {e16133} (\bibinfo {year}
  {2016})}\BibitemShut {NoStop}%
\bibitem [{\citenamefont {Bellouard}\ \emph {et~al.}(2008)\citenamefont
  {Bellouard}, \citenamefont {Barthel}, \citenamefont {Said}, \citenamefont
  {Dugan},\ and\ \citenamefont {Bado}}]{bellouard2008}%
  \BibitemOpen
  \bibfield  {author} {\bibinfo {author} {\bibfnamefont {Y.}~\bibnamefont
  {Bellouard}}, \bibinfo {author} {\bibfnamefont {E.}~\bibnamefont {Barthel}},
  \bibinfo {author} {\bibfnamefont {A.~A.}\ \bibnamefont {Said}}, \bibinfo
  {author} {\bibfnamefont {M.}~\bibnamefont {Dugan}}, \ and\ \bibinfo {author}
  {\bibfnamefont {P.}~\bibnamefont {Bado}},\ }\bibfield  {title} {\emph
  {\bibinfo {title} {Scanning thermal microscopy and {R}aman analysis of bulk
  fused silica exposed to low-energy femtosecond laser pulses},\ }}\href
  {\doibase 10.1364/OE.16.019520} {\bibfield  {journal} {\bibinfo  {journal}
  {Opt. Express}\ }\textbf {\bibinfo {volume} {16}},\ \bibinfo {pages} {19520}
  (\bibinfo {year} {2008})}\BibitemShut {NoStop}%
\bibitem [{\citenamefont {Shcheblanov}\ and\ \citenamefont
  {Povarnitsyn}(2016)}]{shcheblanov2016}%
  \BibitemOpen
  \bibfield  {author} {\bibinfo {author} {\bibfnamefont {N.~S.}\ \bibnamefont
  {Shcheblanov}}\ and\ \bibinfo {author} {\bibfnamefont {M.~E.}\ \bibnamefont
  {Povarnitsyn}},\ }\bibfield  {title} {\emph {\bibinfo {title} {Bond-breaking
  mechanism of vitreous silica densification by {IR} femtosecond laser
  pulses},\ }}\href {\doibase 10.1016/j.jnoncrysol.2015.07.035} {\bibfield
  {journal} {\bibinfo  {journal} {EPL (Europhys. Lett.)}\ }\textbf {\bibinfo
  {volume} {114}},\ \bibinfo {pages} {26004} (\bibinfo {year}
  {2016})}\BibitemShut {NoStop}%
\bibitem [{\citenamefont {Zoubir}\ \emph {et~al.}(2006)\citenamefont {Zoubir},
  \citenamefont {Rivero}, \citenamefont {Grodsky}, \citenamefont {Richardson},
  \citenamefont {Richardson}, \citenamefont {Cardinal},\ and\ \citenamefont
  {Couzi}}]{zoubir2006}%
  \BibitemOpen
  \bibfield  {author} {\bibinfo {author} {\bibfnamefont {A.}~\bibnamefont
  {Zoubir}}, \bibinfo {author} {\bibfnamefont {C.}~\bibnamefont {Rivero}},
  \bibinfo {author} {\bibfnamefont {R.}~\bibnamefont {Grodsky}}, \bibinfo
  {author} {\bibfnamefont {K.}~\bibnamefont {Richardson}}, \bibinfo {author}
  {\bibfnamefont {M.}~\bibnamefont {Richardson}}, \bibinfo {author}
  {\bibfnamefont {T.}~\bibnamefont {Cardinal}}, \ and\ \bibinfo {author}
  {\bibfnamefont {M.}~\bibnamefont {Couzi}},\ }\bibfield  {title} {\emph
  {\bibinfo {title} {Laser-induced defects in fused silica by femtosecond {IR}
  irradiation},\ }}\href {\doibase 10.1103/PhysRevB.73.224117} {\bibfield
  {journal} {\bibinfo  {journal} {Phys. Rev. B}\ }\textbf {\bibinfo {volume}
  {73}},\ \bibinfo {pages} {224117} (\bibinfo {year} {2006})}\BibitemShut
  {NoStop}%
\bibitem [{\citenamefont {Shimotsuma}\ \emph {et~al.}(2011)\citenamefont
  {Shimotsuma}, \citenamefont {Sakakura},\ and\ \citenamefont
  {Miura}}]{shimotsuma2011}%
  \BibitemOpen
  \bibfield  {author} {\bibinfo {author} {\bibfnamefont {Y.}~\bibnamefont
  {Shimotsuma}}, \bibinfo {author} {\bibfnamefont {M.}~\bibnamefont
  {Sakakura}}, \ and\ \bibinfo {author} {\bibfnamefont {K.}~\bibnamefont
  {Miura}},\ }\bibfield  {title} {\emph {\bibinfo {title} {Manipulation of
  optical anisotropy in silica glass},\ }}\href {\doibase 10.1364/OME.1.000803}
  {\bibfield  {journal} {\bibinfo  {journal} {Opt. Mater. Express}\ }\textbf
  {\bibinfo {volume} {1}},\ \bibinfo {pages} {803} (\bibinfo {year}
  {2011})}\BibitemShut {NoStop}%
\bibitem [{\citenamefont {Chan}\ \emph {et~al.}(2001)\citenamefont {Chan},
  \citenamefont {Huser}, \citenamefont {Risbud},\ and\ \citenamefont
  {Krol}}]{Chan2001}%
  \BibitemOpen
  \bibfield  {author} {\bibinfo {author} {\bibfnamefont {J.~W.}\ \bibnamefont
  {Chan}}, \bibinfo {author} {\bibfnamefont {T.}~\bibnamefont {Huser}},
  \bibinfo {author} {\bibfnamefont {S.}~\bibnamefont {Risbud}}, \ and\ \bibinfo
  {author} {\bibfnamefont {D.~M.}\ \bibnamefont {Krol}},\ }\bibfield  {title}
  {\emph {\bibinfo {title} {Structural changes in fused silica after exposure
  to focused femtosecond laser pulses},\ }}\href {\doibase
  10.1364/OL.26.001726} {\bibfield  {journal} {\bibinfo  {journal} {Opt.
  Lett.}\ }\textbf {\bibinfo {volume} {26}},\ \bibinfo {pages} {1726} (\bibinfo
  {year} {2001})}\BibitemShut {NoStop}%
\bibitem [{\citenamefont {Ponader}\ \emph {et~al.}(2008)\citenamefont
  {Ponader}, \citenamefont {Schroeder},\ and\ \citenamefont
  {Streltsov}}]{streltsov2008}%
  \BibitemOpen
  \bibfield  {author} {\bibinfo {author} {\bibfnamefont {C.~W.}\ \bibnamefont
  {Ponader}}, \bibinfo {author} {\bibfnamefont {J.~F.}\ \bibnamefont
  {Schroeder}}, \ and\ \bibinfo {author} {\bibfnamefont {A.~M.}\ \bibnamefont
  {Streltsov}},\ }\bibfield  {title} {\emph {\bibinfo {title} {Origin of the
  refractive-index increase in laser-written waveguides in glasses},\ }}\href
  {\doibase 10.1063/1.2888561} {\bibfield  {journal} {\bibinfo  {journal} {J.
  Appl. Phys.}\ }\textbf {\bibinfo {volume} {103}},\ \bibinfo {pages} {063516}
  (\bibinfo {year} {2008})}\BibitemShut {NoStop}%
\bibitem [{\citenamefont {Mishchik}\ \emph {et~al.}(2013)\citenamefont
  {Mishchik}, \citenamefont {D'Amico}, \citenamefont {Velpula}, \citenamefont
  {Mauclair}, \citenamefont {Boukenter}, \citenamefont {Ouerdane},\ and\
  \citenamefont {Stoian}}]{mishchik2013}%
  \BibitemOpen
  \bibfield  {author} {\bibinfo {author} {\bibfnamefont {K.}~\bibnamefont
  {Mishchik}}, \bibinfo {author} {\bibfnamefont {C.}~\bibnamefont {D'Amico}},
  \bibinfo {author} {\bibfnamefont {P.~K.}\ \bibnamefont {Velpula}}, \bibinfo
  {author} {\bibfnamefont {C.}~\bibnamefont {Mauclair}}, \bibinfo {author}
  {\bibfnamefont {A.}~\bibnamefont {Boukenter}}, \bibinfo {author}
  {\bibfnamefont {Y.}~\bibnamefont {Ouerdane}}, \ and\ \bibinfo {author}
  {\bibfnamefont {R.}~\bibnamefont {Stoian}},\ }\bibfield  {title} {\emph
  {\bibinfo {title} {Ultrafast laser induced electronic and structural
  modifications in bulk fused silica},\ }}\href {\doibase 10.1063/1.4822313}
  {\bibfield  {journal} {\bibinfo  {journal} {J. Appl. Phys.}\ }\textbf
  {\bibinfo {volume} {114}},\ \bibinfo {pages} {133502} (\bibinfo {year}
  {2013})}\BibitemShut {NoStop}%
\bibitem [{\citenamefont {Stoian}\ \emph {et~al.}(2013)\citenamefont {Stoian},
  \citenamefont {Mishchik}, \citenamefont {Cheng}, \citenamefont {Mauclair},
  \citenamefont {D'Amico}, \citenamefont {Colombier},\ and\ \citenamefont
  {Zamfirescu}}]{stoian2013}%
  \BibitemOpen
  \bibfield  {author} {\bibinfo {author} {\bibfnamefont {R.}~\bibnamefont
  {Stoian}}, \bibinfo {author} {\bibfnamefont {K.}~\bibnamefont {Mishchik}},
  \bibinfo {author} {\bibfnamefont {G.}~\bibnamefont {Cheng}}, \bibinfo
  {author} {\bibfnamefont {C.}~\bibnamefont {Mauclair}}, \bibinfo {author}
  {\bibfnamefont {C.}~\bibnamefont {D'Amico}}, \bibinfo {author} {\bibfnamefont
  {J.-P.}\ \bibnamefont {Colombier}}, \ and\ \bibinfo {author} {\bibfnamefont
  {M.}~\bibnamefont {Zamfirescu}},\ }\bibfield  {title} {\emph {\bibinfo
  {title} {Investigation and control of ultrafast laser-induced isotropic and
  anisotropic nanoscale-modulated index patterns in bulk fused silica},\
  }}\href {\doibase 10.1364/OME.3.001755} {\bibfield  {journal} {\bibinfo
  {journal} {Opt. Mat. Express}\ }\textbf {\bibinfo {volume} {3}},\
  \bibinfo {pages} {1755} (\bibinfo {year} {2013})}\BibitemShut {NoStop}%
\bibitem [{\citenamefont {Bellouard}\ \emph {et~al.}(2016)\citenamefont
  {Bellouard}, \citenamefont {Champion}, \citenamefont {McMillen},
  \citenamefont {Mukherjee}, \citenamefont {Thomson}, \citenamefont
  {P{\'e}pin}, \citenamefont {Gillet},\ and\ \citenamefont
  {Cheng}}]{bellouard2016}%
  \BibitemOpen
  \bibfield  {author} {\bibinfo {author} {\bibfnamefont {Y.}~\bibnamefont
  {Bellouard}}, \bibinfo {author} {\bibfnamefont {A.}~\bibnamefont {Champion}},
  \bibinfo {author} {\bibfnamefont {B.}~\bibnamefont {McMillen}}, \bibinfo
  {author} {\bibfnamefont {S.}~\bibnamefont {Mukherjee}}, \bibinfo {author}
  {\bibfnamefont {R.~R.}\ \bibnamefont {Thomson}}, \bibinfo {author}
  {\bibfnamefont {C.}~\bibnamefont {P{\'e}pin}}, \bibinfo {author}
  {\bibfnamefont {P.}~\bibnamefont {Gillet}}, \ and\ \bibinfo {author}
  {\bibfnamefont {Y.}~\bibnamefont {Cheng}},\ }\bibfield  {title} {\emph
  {\bibinfo {title} {Stress-state manipulation in fused silica via femtosecond
  laser irradiation},\ }}\href {\doibase 10.1364/OPTICA.3.001285} {\bibfield
  {journal} {\bibinfo  {journal} {Optica}\ }\textbf {\bibinfo {volume} {3}},\
  \bibinfo {pages} {1285} (\bibinfo {year} {2016})}\BibitemShut {NoStop}%
\bibitem [{\citenamefont {Stoian}\ \emph {et~al.}(2016)\citenamefont {Stoian},
  \citenamefont {D'Amico}, \citenamefont {Bhuyan},\ and\ \citenamefont
  {Cheng}}]{stoian2016}%
  \BibitemOpen
  \bibfield  {author} {\bibinfo {author} {\bibfnamefont {R.}~\bibnamefont
  {Stoian}}, \bibinfo {author} {\bibfnamefont {C.}~\bibnamefont {D'Amico}},
  \bibinfo {author} {\bibfnamefont {M.}~\bibnamefont {Bhuyan}}, \ and\ \bibinfo
  {author} {\bibfnamefont {G.}~\bibnamefont {Cheng}},\ }\bibfield  {title}
  {\emph {\bibinfo {title} {Ultrafast laser photoinscription of large-mode-area
  waveguiding structures in bulk dielectrics: {I}nvited paper for the section:
  {H}ot topics in ultrafast lasers},\ }}\href {\doibase
  10.1016/j.optlastec.2015.11.025} {\bibfield  {journal} {\bibinfo  {journal}
  {Opt. Laser Technol.}\ }\textbf {\bibinfo {volume} {80}},\ \bibinfo {pages}
  {98} (\bibinfo {year} {2016})}\BibitemShut {NoStop}%
\bibitem [{\citenamefont {Hernandez-Rueda}\ \emph {et~al.}(2017)\citenamefont
  {Hernandez-Rueda}, \citenamefont {Clarijs}, \citenamefont {van Oosten},\ and\
  \citenamefont {Krol}}]{Krol2017}%
  \BibitemOpen
  \bibfield  {author} {\bibinfo {author} {\bibfnamefont {J.}~\bibnamefont
  {Hernandez-Rueda}}, \bibinfo {author} {\bibfnamefont {J.}~\bibnamefont
  {Clarijs}}, \bibinfo {author} {\bibfnamefont {D.}~\bibnamefont {van Oosten}},
  \ and\ \bibinfo {author} {\bibfnamefont {D.~M.}\ \bibnamefont {Krol}},\
  }\bibfield  {title} {\emph {\bibinfo {title} {The influence of femtosecond
  laser wavelength on waveguide fabrication inside fused silica},\ }}\href
  {\doibase 10.1063/1.4981124} {\bibfield  {journal} {\bibinfo  {journal}
  {Appl. Phys. Lett.}\ }\textbf {\bibinfo {volume} {110}},\ \bibinfo {pages}
  {161109} (\bibinfo {year} {2017})}\BibitemShut {NoStop}%
\bibitem [{\citenamefont {Barrio}\ \emph {et~al.}(1993)\citenamefont {Barrio},
  \citenamefont {Galeener}, \citenamefont {Martinez},\ and\ \citenamefont
  {Elliott}}]{galeener93}%
  \BibitemOpen
  \bibfield  {author} {\bibinfo {author} {\bibfnamefont {R.~A.}\ \bibnamefont
  {Barrio}}, \bibinfo {author} {\bibfnamefont {F.~L.}\ \bibnamefont
  {Galeener}}, \bibinfo {author} {\bibfnamefont {E.}~\bibnamefont {Martinez}},
  \ and\ \bibinfo {author} {\bibfnamefont {R.~J.}\ \bibnamefont {Elliott}},\
  }\bibfield  {title} {\emph {\bibinfo {title} {Regular ring dynamics in
  {AX}$_2$ tetrahedral glasses},\ }}\href {\doibase 10.1103/PhysRevB.48.15672}
  {\bibfield  {journal} {\bibinfo  {journal} {Phys. Rev. B}\ }\textbf {\bibinfo
  {volume} {48}},\ \bibinfo {pages} {15672} (\bibinfo {year}
  {1993})}\BibitemShut {NoStop}%
\bibitem [{\citenamefont {Pasquarello}\ and\ \citenamefont
  {Car}(1998)}]{alfprl98}%
  \BibitemOpen
  \bibfield  {author} {\bibinfo {author} {\bibfnamefont {A.}~\bibnamefont
  {Pasquarello}}\ and\ \bibinfo {author} {\bibfnamefont {R.}~\bibnamefont
  {Car}},\ }\bibfield  {title} {\emph {\bibinfo {title} {Identification of
  {R}aman defect lines as signatures of ring structures in vitreous silica},\
  }}\href {\doibase 10.1103/PhysRevLett.80.5145} {\bibfield  {journal}
  {\bibinfo  {journal} {Phys. Rev. Lett.}\ }\textbf {\bibinfo {volume} {80}},\
  \bibinfo {pages} {5145} (\bibinfo {year} {1998})}\BibitemShut {NoStop}%
\bibitem [{\citenamefont {Huang}\ \emph {et~al.}(2013)\citenamefont {Huang},
  \citenamefont {Kurasch}, \citenamefont {Alden}, \citenamefont {Shekhawat},
  \citenamefont {Alemi}, \citenamefont {McEuen}, \citenamefont {Sethna},
  \citenamefont {Kaiser},\ and\ \citenamefont {Muller}}]{huang2013}%
  \BibitemOpen
  \bibfield  {author} {\bibinfo {author} {\bibfnamefont {P.~Y.}\ \bibnamefont
  {Huang}}, \bibinfo {author} {\bibfnamefont {S.}~\bibnamefont {Kurasch}},
  \bibinfo {author} {\bibfnamefont {J.~S.}\ \bibnamefont {Alden}}, \bibinfo
  {author} {\bibfnamefont {A.}~\bibnamefont {Shekhawat}}, \bibinfo {author}
  {\bibfnamefont {A.~A.}\ \bibnamefont {Alemi}}, \bibinfo {author}
  {\bibfnamefont {P.~L.}\ \bibnamefont {McEuen}}, \bibinfo {author}
  {\bibfnamefont {J.~P.}\ \bibnamefont {Sethna}}, \bibinfo {author}
  {\bibfnamefont {U.}~\bibnamefont {Kaiser}}, \ and\ \bibinfo {author}
  {\bibfnamefont {D.~A.}\ \bibnamefont {Muller}},\ }\bibfield  {title} {\emph
  {\bibinfo {title} {Imaging atomic rearrangements in two-dimensional silica
  glass: Watching silica's dance},\ }}\href {\doibase 10.1126/science.1242248}
  {\bibfield  {journal} {\bibinfo  {journal} {Science}\ }\textbf {\bibinfo
  {volume} {342}},\ \bibinfo {pages} {224} (\bibinfo {year}
  {2013})}\BibitemShut {NoStop}%
\bibitem [{\citenamefont {Bressel}\ \emph {et~al.}(2011)\citenamefont
  {Bressel}, \citenamefont {de~Ligny}, \citenamefont {Sonneville},
  \citenamefont {Martinez}, \citenamefont {Mizeikis}, \citenamefont
  {Buividas},\ and\ \citenamefont {Juodkazis}}]{bressel2011}%
  \BibitemOpen
  \bibfield  {author} {\bibinfo {author} {\bibfnamefont {M.}~\bibnamefont
  {Bressel}}, \bibinfo {author} {\bibfnamefont {D.}~\bibnamefont {de~Ligny}},
  \bibinfo {author} {\bibfnamefont {C.}~\bibnamefont {Sonneville}}, \bibinfo
  {author} {\bibfnamefont {V.}~\bibnamefont {Martinez}}, \bibinfo {author}
  {\bibfnamefont {V.}~\bibnamefont {Mizeikis}}, \bibinfo {author}
  {\bibfnamefont {R.}~\bibnamefont {Buividas}}, \ and\ \bibinfo {author}
  {\bibfnamefont {S.}~\bibnamefont {Juodkazis}},\ }\bibfield  {title} {\emph
  {\bibinfo {title} {Femtosecond laser induced density changes in {G}e{O}$_2$
  and {S}i{O}$_2$ glasses: fictive temperature effect},\ }}\href {\doibase
  10.1364/OME.1.000605} {\bibfield  {journal} {\bibinfo  {journal} {Opt. Mat.
  Express}\ }\textbf {\bibinfo {volume} {1}},\ \bibinfo {pages} {605} (\bibinfo
  {year} {2011})}\BibitemShut {NoStop}%
\bibitem [{\citenamefont {Deschamps}\ \emph {et~al.}(2014)\citenamefont
  {Deschamps}, \citenamefont {Margueritat}, \citenamefont {Martinet},
  \citenamefont {Mermet},\ and\ \citenamefont {Champagnon}}]{deschamps2014}%
  \BibitemOpen
  \bibfield  {author} {\bibinfo {author} {\bibfnamefont {T.}~\bibnamefont
  {Deschamps}}, \bibinfo {author} {\bibfnamefont {J.}~\bibnamefont
  {Margueritat}}, \bibinfo {author} {\bibfnamefont {C.}~\bibnamefont
  {Martinet}}, \bibinfo {author} {\bibfnamefont {A.}~\bibnamefont {Mermet}}, \
  and\ \bibinfo {author} {\bibfnamefont {B.}~\bibnamefont {Champagnon}},\
  }\bibfield  {title} {\emph {\bibinfo {title} {Elastic moduli of permanently
  densified silica glasses},\ }}\href {\doibase 10.1038/srep07193} {\bibfield
  {journal} {\bibinfo  {journal} {Sci. Rep.}\ }\textbf {\bibinfo {volume}
  {4}},\ \bibinfo {pages} {7193} (\bibinfo {year} {2014})}\BibitemShut
  {NoStop}%
\bibitem [{\citenamefont {Martinet}\ \emph {et~al.}(2015)\citenamefont
  {Martinet}, \citenamefont {Kassir-Bodon}, \citenamefont {Deschamps},
  \citenamefont {Cornet}, \citenamefont {{L}e Floch}, \citenamefont
  {Martinez},\ and\ \citenamefont {Champagnon}}]{martinet2015}%
  \BibitemOpen
  \bibfield  {author} {\bibinfo {author} {\bibfnamefont {C.}~\bibnamefont
  {Martinet}}, \bibinfo {author} {\bibfnamefont {A.}~\bibnamefont
  {Kassir-Bodon}}, \bibinfo {author} {\bibfnamefont {T.}~\bibnamefont
  {Deschamps}}, \bibinfo {author} {\bibfnamefont {A.}~\bibnamefont {Cornet}},
  \bibinfo {author} {\bibfnamefont {S.}~\bibnamefont {{L}e Floch}}, \bibinfo
  {author} {\bibfnamefont {V.}~\bibnamefont {Martinez}}, \ and\ \bibinfo
  {author} {\bibfnamefont {B.}~\bibnamefont {Champagnon}},\ }\bibfield  {title}
  {\emph {\bibinfo {title} {Permanently densified {S}i{O}$_2$ glasses: a
  structural approach},\ }}\href {\doibase 10.1088/0953-8984/27/32/325401}
  {\bibfield  {journal} {\bibinfo  {journal} {J. Phys.: Condens. Mat.}\
  }\textbf {\bibinfo {volume} {27}},\ \bibinfo {pages} {325401} (\bibinfo
  {year} {2015})}\BibitemShut {NoStop}%
\bibitem [{\citenamefont {Cornet}\ \emph {et~al.}(2017)\citenamefont {Cornet},
  \citenamefont {Martinez}, \citenamefont {de~Ligny}, \citenamefont
  {Champagnon},\ and\ \citenamefont {Martinet}}]{cornet2017}%
  \BibitemOpen
  \bibfield  {author} {\bibinfo {author} {\bibfnamefont {A.}~\bibnamefont
  {Cornet}}, \bibinfo {author} {\bibfnamefont {V.}~\bibnamefont {Martinez}},
  \bibinfo {author} {\bibfnamefont {D.}~\bibnamefont {de~Ligny}}, \bibinfo
  {author} {\bibfnamefont {B.}~\bibnamefont {Champagnon}}, \ and\ \bibinfo
  {author} {\bibfnamefont {C.}~\bibnamefont {Martinet}},\ }\bibfield  {title}
  {\emph {\bibinfo {title} {Relaxation processes of densified silica glass},\
  }}\href {\doibase 10.1063/1.4977036} {\bibfield  {journal} {\bibinfo
  {journal} {J. Chem. Phys.}\ }\textbf {\bibinfo {volume} {146}},\ \bibinfo
  {pages} {094504} (\bibinfo {year} {2017})}\BibitemShut {NoStop}%
\bibitem [{\citenamefont {Shcheblanov}\ \emph {et~al.}(2015)\citenamefont
  {Shcheblanov}, \citenamefont {Mantisi}, \citenamefont {Umari},\ and\
  \citenamefont {Tanguy}}]{shcheblanov2015}%
  \BibitemOpen
  \bibfield  {author} {\bibinfo {author} {\bibfnamefont {N.~S.}\ \bibnamefont
  {Shcheblanov}}, \bibinfo {author} {\bibfnamefont {B.}~\bibnamefont
  {Mantisi}}, \bibinfo {author} {\bibfnamefont {P.}~\bibnamefont {Umari}}, \
  and\ \bibinfo {author} {\bibfnamefont {A.}~\bibnamefont {Tanguy}},\
  }\bibfield  {title} {\emph {\bibinfo {title} {Detailed analysis of plastic
  shear in the {R}aman spectra of {S}i{O}$_2$ glass},\ }}\href {\doibase
  10.1016/j.jnoncrysol.2015.07.035} {\bibfield  {journal} {\bibinfo  {journal}
  {J. Non-Cryst. Solids}\ }\textbf {\bibinfo {volume} {428}},\ \bibinfo {pages}
  {6} (\bibinfo {year} {2015})}\BibitemShut {NoStop}%
\bibitem [{\citenamefont {Sen}\ and\ \citenamefont
  {Thorpe}(1977)}]{thorpe1977}%
  \BibitemOpen
  \bibfield  {author} {\bibinfo {author} {\bibfnamefont {P.~N.}\ \bibnamefont
  {Sen}}\ and\ \bibinfo {author} {\bibfnamefont {M.~F.}\ \bibnamefont
  {Thorpe}},\ }\bibfield  {title} {\emph {\bibinfo {title} {Phonons in {AX}$_2$
  glasses: {F}rom molecular to band-like modes},\ }}\href {\doibase
  doi.org/10.1103/PhysRevB.15.4030} {\bibfield  {journal} {\bibinfo  {journal}
  {Phys. Rev. B}\ }\textbf {\bibinfo {volume} {15}},\ \bibinfo {pages} {4030}
  (\bibinfo {year} {1977})}\BibitemShut {NoStop}%
\bibitem [{\citenamefont {Hehlen}(2010)}]{hehlen2010}%
  \BibitemOpen
  \bibfield  {author} {\bibinfo {author} {\bibfnamefont {B.}~\bibnamefont
  {Hehlen}},\ }\bibfield  {title} {\emph {\bibinfo {title} {Inter-tetrahedra
  bond angle of permanently densified silicas extracted from their {R}aman
  spectra},\ }}\href {\doibase 10.1088/0953-8984/22/2/025401} {\bibfield
  {journal} {\bibinfo  {journal} {J. Phys.: Condens. Mat.}\ }\textbf {\bibinfo
  {volume} {22}},\ \bibinfo {pages} {025401} (\bibinfo {year}
  {2010})}\BibitemShut {NoStop}%
\bibitem [{\citenamefont {Weigel}\ \emph {et~al.}(2016)\citenamefont {Weigel},
  \citenamefont {Foret}, \citenamefont {Hehlen}, \citenamefont {Kint},
  \citenamefont {Clement}, \citenamefont {Polian}, \citenamefont {Vacher},\
  and\ \citenamefont {Ruffl{\'e}}}]{weigel2016}%
  \BibitemOpen
  \bibfield  {author} {\bibinfo {author} {\bibfnamefont {C.}~\bibnamefont
  {Weigel}}, \bibinfo {author} {\bibfnamefont {M.}~\bibnamefont {Foret}},
  \bibinfo {author} {\bibfnamefont {B.}~\bibnamefont {Hehlen}}, \bibinfo
  {author} {\bibfnamefont {M.}~\bibnamefont {Kint}}, \bibinfo {author}
  {\bibfnamefont {S.}~\bibnamefont {Clement}}, \bibinfo {author} {\bibfnamefont
  {A.}~\bibnamefont {Polian}}, \bibinfo {author} {\bibfnamefont
  {R.}~\bibnamefont {Vacher}}, \ and\ \bibinfo {author} {\bibfnamefont
  {B.}~\bibnamefont {Ruffl{\'e}}},\ }\bibfield  {title} {\emph {\bibinfo
  {title} {Polarized {R}aman spectroscopy of $v$-{S}i{O}$_2$ under rare-gas
  compression},\ }}\href {\doibase 10.1103/PhysRevB.93.224303} {\bibfield
  {journal} {\bibinfo  {journal} {Phys. Rev. B}\ }\textbf {\bibinfo {volume}
  {93}},\ \bibinfo {pages} {224303} (\bibinfo {year} {2016})}\BibitemShut
  {NoStop}%
\bibitem [{\citenamefont {Eaton}\ \emph {et~al.}(2008)\citenamefont {Eaton},
  \citenamefont {Zhang}, \citenamefont {Ng}, \citenamefont {Li}, \citenamefont
  {Chen}, \citenamefont {Ho},\ and\ \citenamefont {Herman}}]{eaton2008}%
  \BibitemOpen
  \bibfield  {author} {\bibinfo {author} {\bibfnamefont {S.~M.}\ \bibnamefont
  {Eaton}}, \bibinfo {author} {\bibfnamefont {H.}~\bibnamefont {Zhang}},
  \bibinfo {author} {\bibfnamefont {M.~L.}\ \bibnamefont {Ng}}, \bibinfo
  {author} {\bibfnamefont {J.}~\bibnamefont {Li}}, \bibinfo {author}
  {\bibfnamefont {W.-J.}\ \bibnamefont {Chen}}, \bibinfo {author}
  {\bibfnamefont {S.}~\bibnamefont {Ho}}, \ and\ \bibinfo {author}
  {\bibfnamefont {P.~R.}\ \bibnamefont {Herman}},\ }\bibfield  {title} {\emph
  {\bibinfo {title} {Transition from thermal diffusion to heat accumulation in
  high repetition rate femtosecond laser writing of buried optical
  waveguides},\ }}\href {\doibase 10.1364/OE.16.009443} {\bibfield  {journal}
  {\bibinfo  {journal} {Opt. Express}\ }\textbf {\bibinfo {volume} {16}},\
  \bibinfo {pages} {9443} (\bibinfo {year} {2008})}\BibitemShut {NoStop}%
\bibitem [{\citenamefont {Sugioka}\ and\ \citenamefont
  {Cheng}(2014)}]{sugioka2014}%
  \BibitemOpen
  \bibfield  {author} {\bibinfo {author} {\bibfnamefont {K.}~\bibnamefont
  {Sugioka}}\ and\ \bibinfo {author} {\bibfnamefont {Y.}~\bibnamefont
  {Cheng}},\ }\bibfield  {title} {\emph {\bibinfo {title} {Ultrafast
  lasers--reliable tools for advanced materials processing},\ }}\href {\doibase
  10.1038/lsa.2014.30} {\bibfield  {journal} {\bibinfo  {journal} {Light: Sci.
  Appl.}\ }\textbf {\bibinfo {volume} {3}},\ \bibinfo {pages} {e149} (\bibinfo
  {year} {2014})}\BibitemShut {NoStop}%
\bibitem [{\citenamefont {Zachariasen}(1932)}]{zachariasen1932}%
  \BibitemOpen
  \bibfield  {author} {\bibinfo {author} {\bibfnamefont {W.~H.}\ \bibnamefont
  {Zachariasen}},\ }\bibfield  {title} {\emph {\bibinfo {title} {The atomic
  arrangement in glass},\ }}\href {\doibase 10.1021/ja01349a006} {\bibfield
  {journal} {\bibinfo  {journal} {J. Am. Chem. Soc.}\ }\textbf {\bibinfo
  {volume} {54}},\ \bibinfo {pages} {3841} (\bibinfo {year}
  {1932})}\BibitemShut {NoStop}%
\bibitem [{\citenamefont {Mantisi}\ \emph {et~al.}(2012)\citenamefont
  {Mantisi}, \citenamefont {Tanguy}, \citenamefont {Kermouche},\ and\
  \citenamefont {Barthel}}]{boris2012}%
  \BibitemOpen
  \bibfield  {author} {\bibinfo {author} {\bibfnamefont {B.}~\bibnamefont
  {Mantisi}}, \bibinfo {author} {\bibfnamefont {A.}~\bibnamefont {Tanguy}},
  \bibinfo {author} {\bibfnamefont {G.}~\bibnamefont {Kermouche}}, \ and\
  \bibinfo {author} {\bibfnamefont {E.}~\bibnamefont {Barthel}},\ }\bibfield
  {title} {\emph {\bibinfo {title} {Atomistic response of a model silica glass
  under shear and pressure},\ }}\href {\doibase 10.1140/epjb/e2012-30317-6}
  {\bibfield  {journal} {\bibinfo  {journal} {Eur. Phys. J. B}\ }\textbf
  {\bibinfo {volume} {85}},\ \bibinfo {pages} {1} (\bibinfo {year}
  {2012})}\BibitemShut {NoStop}%
\bibitem [{\citenamefont {Plimpton}(1995)}]{lammps}%
  \BibitemOpen
  \bibfield  {author} {\bibinfo {author} {\bibfnamefont {S.}~\bibnamefont
  {Plimpton}},\ }\bibfield  {title} {\emph {\bibinfo {title} {Fast parallel
  algorithms for short-range molecular dynamics},\ }}\href {\doibase
  10.1006/jcph.1995.1039} {\bibfield  {journal} {\bibinfo  {journal} {J.
  Comput. Phys.}\ }\textbf {\bibinfo {volume} {117}},\ \bibinfo {pages} {1}
  (\bibinfo {year} {1995})}\BibitemShut {NoStop}%
\bibitem [{\citenamefont {Binder}\ and\ \citenamefont
  {Kob}(2011)}]{binder2011}%
  \BibitemOpen
  \bibfield  {author} {\bibinfo {author} {\bibfnamefont {K.}~\bibnamefont
  {Binder}}\ and\ \bibinfo {author} {\bibfnamefont {W.}~\bibnamefont {Kob}},\
  }\href@noop {} {\emph {\bibinfo {title} {Glassy materials and disordered
  solids: An introduction to their statistical mechanics}}}\ (\bibinfo
  {publisher} {World Scientific, Singapore},\ \bibinfo {year}
  {2011})\BibitemShut {NoStop}%
\bibitem [{\citenamefont {van Beest}\ \emph {et~al.}(1990)\citenamefont {van
  Beest}, \citenamefont {Kramer},\ and\ \citenamefont {van Santen}}]{bks1990}%
  \BibitemOpen
  \bibfield  {author} {\bibinfo {author} {\bibfnamefont {B.~W.~H.}\
  \bibnamefont {van Beest}}, \bibinfo {author} {\bibfnamefont {G.~J.}\
  \bibnamefont {Kramer}}, \ and\ \bibinfo {author} {\bibfnamefont {R.~A.}\
  \bibnamefont {van Santen}},\ }\bibfield  {title} {\emph {\bibinfo {title}
  {Force fields for silicas and aluminophosphates based on \emph{ab initio}
  calculations},\ }}\href {\doibase 10.1103/PhysRevLett.64.1955} {\bibfield
  {journal} {\bibinfo  {journal} {Phys. Rev. Lett.}\ }\textbf {\bibinfo
  {volume} {64}},\ \bibinfo {pages} {1955} (\bibinfo {year}
  {1990})}\BibitemShut {NoStop}%
\bibitem [{\citenamefont {Carr\'{e}}\ \emph {et~al.}(2007)\citenamefont
  {Carr\'{e}}, \citenamefont {Berthier}, \citenamefont {Horbach}, \citenamefont
  {Ispas},\ and\ \citenamefont {Kob}}]{carre2007}%
  \BibitemOpen
  \bibfield  {author} {\bibinfo {author} {\bibfnamefont {A.}~\bibnamefont
  {Carr\'{e}}}, \bibinfo {author} {\bibfnamefont {L.}~\bibnamefont {Berthier}},
  \bibinfo {author} {\bibfnamefont {J.}~\bibnamefont {Horbach}}, \bibinfo
  {author} {\bibfnamefont {S.}~\bibnamefont {Ispas}}, \ and\ \bibinfo {author}
  {\bibfnamefont {W.}~\bibnamefont {Kob}},\ }\bibfield  {title} {\emph
  {\bibinfo {title} {Amorphous silica modeled with truncated and screened
  {C}oulomb interactions: A molecular dynamics simulation study},\ }}\href
  {\doibase 10.1063/1.2777136} {\bibfield  {journal} {\bibinfo  {journal} {J.
  Chem. Phys.}\ }\textbf {\bibinfo {volume} {127}},\ \bibinfo {pages} {114512}
  (\bibinfo {year} {2007})}\BibitemShut {NoStop}%
\bibitem [{\citenamefont {Cardona}\ and\ \citenamefont
  {Guentherodt}(1982)}]{cardona1982}%
  \BibitemOpen
  \bibfield  {author} {\bibinfo {author} {\bibfnamefont {M.}~\bibnamefont
  {Cardona}}\ and\ \bibinfo {author} {\bibfnamefont {G.}~\bibnamefont
  {Guentherodt}},\ }\href@noop {} {\emph {\bibinfo {title} {Light Scattering in
  Solids II}}}\ (\bibinfo  {publisher} {Springer-Verlag},\ \bibinfo {year}
  {1982})\BibitemShut {NoStop}%
\bibitem [{\citenamefont {Bruesch}(1986)}]{bruesch1986}%
  \BibitemOpen
  \bibfield  {author} {\bibinfo {author} {\bibfnamefont {P.}~\bibnamefont
  {Bruesch}},\ }\href@noop {} {\emph {\bibinfo {title} {Phonons: {T}heory and
  {E}xperiments. Vol. 2}}}\ (\bibinfo  {publisher} {Springer, Berlin},\
  \bibinfo {year} {1986})\BibitemShut {NoStop}%
\bibitem [{\citenamefont {Wolkenstein}(1941)}]{wolk1941}%
  \BibitemOpen
  \bibfield  {author} {\bibinfo {author} {\bibfnamefont {M.~W.}\ \bibnamefont
  {Wolkenstein}},\ }\bibfield  {title} {\emph {\bibinfo {title} {Intensities of
  vibrational spectra of molecules},\ }}\href@noop {} {\bibfield  {journal}
  {\bibinfo  {journal} {C. R. Acad. Sci. URSS}\ }\textbf {\bibinfo {volume}
  {30}},\ \bibinfo {pages} {791} (\bibinfo {year} {1941})}\BibitemShut
  {NoStop}%
\bibitem [{\citenamefont {Eliashevich}\ and\ \citenamefont
  {Wolkenstein}(1945)}]{wolk1944}%
  \BibitemOpen
  \bibfield  {author} {\bibinfo {author} {\bibfnamefont {M.}~\bibnamefont
  {Eliashevich}}\ and\ \bibinfo {author} {\bibfnamefont {M.}~\bibnamefont
  {Wolkenstein}},\ }\bibfield  {title} {\emph {\bibinfo {title} {The theory of
  intensities and polarization of vibration spectra of molecules},\
  }}\href@noop {} {\bibfield  {journal} {\bibinfo  {journal} {J. Phys. USSR}\
  }\textbf {\bibinfo {volume} {9}},\ \bibinfo {pages} {101} (\bibinfo {year}
  {1945})}\BibitemShut {NoStop}%
\bibitem [{\citenamefont {Taraskin}\ and\ \citenamefont
  {Elliott}(1997)}]{taraskin97}%
  \BibitemOpen
  \bibfield  {author} {\bibinfo {author} {\bibfnamefont {S.~N.}\ \bibnamefont
  {Taraskin}}\ and\ \bibinfo {author} {\bibfnamefont {S.~R.}\ \bibnamefont
  {Elliott}},\ }\bibfield  {title} {\emph {\bibinfo {title} {Nature of
  vibrational excitations in vitreous silica},\ }}\href {\doibase
  10.1103/PhysRevB.56.8605} {\bibfield  {journal} {\bibinfo  {journal} {Phys.
  Rev. B}\ }\textbf {\bibinfo {volume} {56}},\ \bibinfo {pages} {8605}
  (\bibinfo {year} {1997})}\BibitemShut {NoStop}%
\bibitem [{\citenamefont {Shcheblanov}\ \emph {et~al.}(2016)\citenamefont
  {Shcheblanov}, \citenamefont {Povarnitsyn}, \citenamefont {Taraskin},\ and\
  \citenamefont {Elliott}}]{taraskin2016}%
  \BibitemOpen
  \bibfield  {author} {\bibinfo {author} {\bibfnamefont {N.~S.}\ \bibnamefont
  {Shcheblanov}}, \bibinfo {author} {\bibfnamefont {M.~E.}\ \bibnamefont
  {Povarnitsyn}}, \bibinfo {author} {\bibfnamefont {S.~N.}\ \bibnamefont
  {Taraskin}}, \ and\ \bibinfo {author} {\bibfnamefont {S.~R.}\ \bibnamefont
  {Elliott}},\ }\bibfield  {title} {\emph {\bibinfo {title} {Addendum and
  {E}rratum: {N}ature of vibrational excitations in vitreous silica [{P}hys.
  {R}ev. {B} 56, 8605 (1997)]},\ }}\href {\doibase 10.1103/PhysRevB.94.099903}
  {\bibfield  {journal} {\bibinfo  {journal} {Phys. Rev. B}\ }\textbf {\bibinfo
  {volume} {94}},\ \bibinfo {pages} {099903} (\bibinfo {year}
  {2016})}\BibitemShut {NoStop}%
\bibitem [{\citenamefont {Sarnthein}\ \emph {et~al.}(1997)\citenamefont
  {Sarnthein}, \citenamefont {Pasquarello},\ and\ \citenamefont {Car}}]{alf97}%
  \BibitemOpen
  \bibfield  {author} {\bibinfo {author} {\bibfnamefont {J.}~\bibnamefont
  {Sarnthein}}, \bibinfo {author} {\bibfnamefont {A.}~\bibnamefont
  {Pasquarello}}, \ and\ \bibinfo {author} {\bibfnamefont {R.}~\bibnamefont
  {Car}},\ }\bibfield  {title} {\emph {\bibinfo {title} {Origin of the
  high-frequency doublet in the vibrational spectrum of vitreous {S}i{O}$_2$},\
  }}\href {\doibase 10.1126/science.275.5308.1925} {\bibfield  {journal}
  {\bibinfo  {journal} {Science}\ }\textbf {\bibinfo {volume} {275}},\ \bibinfo
  {pages} {1925} (\bibinfo {year} {1997})}\BibitemShut {NoStop}%
\bibitem [{\citenamefont {Umari}\ and\ \citenamefont
  {Pasquarello}(2003)}]{umari03}%
  \BibitemOpen
  \bibfield  {author} {\bibinfo {author} {\bibfnamefont {P.}~\bibnamefont
  {Umari}}\ and\ \bibinfo {author} {\bibfnamefont {A.}~\bibnamefont
  {Pasquarello}},\ }\bibfield  {title} {\emph {\bibinfo {title}
  {First-principles analysis of the {R}aman spectrum of vitreous silica:
  comparison with the vibrational density of states},\ }}\href {\doibase
  10.1088/0953-8984/15/16/304} {\bibfield  {journal} {\bibinfo  {journal} {J.
  Phys.: Condens. Mat.}\ }\textbf {\bibinfo {volume} {15}},\ \bibinfo {pages}
  {S1547} (\bibinfo {year} {2003})}\BibitemShut {NoStop}%
\bibitem [{\citenamefont {Trease}\ \emph {et~al.}(2017)\citenamefont {Trease},
  \citenamefont {Clark}, \citenamefont {Grandinetti}, \citenamefont
  {Stebbins},\ and\ \citenamefont {Sen}}]{trease2017}%
  \BibitemOpen
  \bibfield  {author} {\bibinfo {author} {\bibfnamefont {N.~M.}\ \bibnamefont
  {Trease}}, \bibinfo {author} {\bibfnamefont {T.~M.}\ \bibnamefont {Clark}},
  \bibinfo {author} {\bibfnamefont {P.~J.}\ \bibnamefont {Grandinetti}},
  \bibinfo {author} {\bibfnamefont {J.~F.}\ \bibnamefont {Stebbins}}, \ and\
  \bibinfo {author} {\bibfnamefont {S.}~\bibnamefont {Sen}},\ }\bibfield
  {title} {\emph {\bibinfo {title} {Bond length-bond angle correlation in
  densified silica---{R}esults from $^{17}${O} {NMR} spectroscopy},\ }}\href
  {\doibase 10.1063/1.4983041} {\bibfield  {journal} {\bibinfo  {journal} {J.
  Chem. Phys.}\ }\textbf {\bibinfo {volume} {146}},\ \bibinfo {pages} {184505}
  (\bibinfo {year} {2017})}\BibitemShut {NoStop}%
\bibitem [{\citenamefont {Umari}\ \emph {et~al.}(2003)\citenamefont {Umari},
  \citenamefont {Gonze},\ and\ \citenamefont {Pasquarello}}]{umari2003}%
  \BibitemOpen
  \bibfield  {author} {\bibinfo {author} {\bibfnamefont {P.}~\bibnamefont
  {Umari}}, \bibinfo {author} {\bibfnamefont {X.}~\bibnamefont {Gonze}}, \ and\
  \bibinfo {author} {\bibfnamefont {A.}~\bibnamefont {Pasquarello}},\
  }\bibfield  {title} {\emph {\bibinfo {title} {Concentration of small ring
  structures in vitreous silica from a first-principles analysis of the {R}aman
  spectrum},\ }}\href {\doibase 10.1103/PhysRevLett.90.027401} {\bibfield
  {journal} {\bibinfo  {journal} {Phys. Rev. Lett.}\ }\textbf {\bibinfo
  {volume} {90}},\ \bibinfo {pages} {027401} (\bibinfo {year}
  {2003})}\BibitemShut {NoStop}%
\bibitem [{\citenamefont {Hiramatsu}\ \emph {et~al.}(1996)\citenamefont
  {Hiramatsu}, \citenamefont {Arai}, \citenamefont {Shibazaki}, \citenamefont
  {Tsunekawa}, \citenamefont {Otomo}, \citenamefont {Hannon}, \citenamefont
  {Bennington}, \citenamefont {Kitamura},\ and\ \citenamefont
  {Onodera}}]{hiramatsu1996}%
  \BibitemOpen
  \bibfield  {author} {\bibinfo {author} {\bibfnamefont {A.}~\bibnamefont
  {Hiramatsu}}, \bibinfo {author} {\bibfnamefont {M.}~\bibnamefont {Arai}},
  \bibinfo {author} {\bibfnamefont {H.}~\bibnamefont {Shibazaki}}, \bibinfo
  {author} {\bibfnamefont {M.}~\bibnamefont {Tsunekawa}}, \bibinfo {author}
  {\bibfnamefont {T.}~\bibnamefont {Otomo}}, \bibinfo {author} {\bibfnamefont
  {A.~C.}\ \bibnamefont {Hannon}}, \bibinfo {author} {\bibfnamefont {S.~M.}\
  \bibnamefont {Bennington}}, \bibinfo {author} {\bibfnamefont
  {N.}~\bibnamefont {Kitamura}}, \ and\ \bibinfo {author} {\bibfnamefont
  {A.}~\bibnamefont {Onodera}},\ }\bibfield  {title} {\emph {\bibinfo {title}
  {Investigation on permanently densified vitreous silica by means of neutron
  scattering},\ }}\href {\doibase 10.1016/0921-4526(95)00722-9} {\bibfield
  {journal} {\bibinfo  {journal} {Physica B}\ }\textbf {\bibinfo {volume}
  {219-220}},\ \bibinfo {pages} {287} (\bibinfo {year} {1996})}\BibitemShut
  {NoStop}%
\bibitem [{\citenamefont {Malfait}\ \emph {et~al.}(2008)\citenamefont
  {Malfait}, \citenamefont {Halter},\ and\ \citenamefont
  {Verel}}]{malfait2008}%
  \BibitemOpen
  \bibfield  {author} {\bibinfo {author} {\bibfnamefont {W.~J.}\ \bibnamefont
  {Malfait}}, \bibinfo {author} {\bibfnamefont {W.~E.}\ \bibnamefont {Halter}},
  \ and\ \bibinfo {author} {\bibfnamefont {R.}~\bibnamefont {Verel}},\
  }\bibfield  {title} {\emph {\bibinfo {title} {$^{29}${S}i {NMR} spectroscopy
  of silica glass: {T}$_1$ relaxation and constraints on the {S}i--{O}--{S}i
  bond angle distribution},\ }}\href {\doibase 10.1016/j.chemgeo.2008.06.048}
  {\bibfield  {journal} {\bibinfo  {journal} {Chem. Geol.}\ }\textbf {\bibinfo
  {volume} {256}},\ \bibinfo {pages} {269} (\bibinfo {year}
  {2008})}\BibitemShut {NoStop}%
\bibitem [{\citenamefont {Geissberger}\ and\ \citenamefont
  {Galeener}(1983)}]{geissberger83}%
  \BibitemOpen
  \bibfield  {author} {\bibinfo {author} {\bibfnamefont {A.}~\bibnamefont
  {Geissberger}}\ and\ \bibinfo {author} {\bibfnamefont {F.~L.}\ \bibnamefont
  {Galeener}},\ }\bibfield  {title} {\emph {\bibinfo {title} {Raman studies of
  vitreous {S}i{O}$_2$ versus fictive temperature},\ }}\href {\doibase
  10.1103/PhysRevB.28.3266} {\bibfield  {journal} {\bibinfo  {journal} {Phys.
  Rev. B}\ }\textbf {\bibinfo {volume} {28}},\ \bibinfo {pages} {3266}
  (\bibinfo {year} {1983})}\BibitemShut {NoStop}%
\bibitem [{\citenamefont {Trave}\ \emph {et~al.}(2002)\citenamefont {Trave},
  \citenamefont {Tangney}, \citenamefont {Scandolo}, \citenamefont
  {Pasquarello},\ and\ \citenamefont {Car}}]{trave2002}%
  \BibitemOpen
  \bibfield  {author} {\bibinfo {author} {\bibfnamefont {A.}~\bibnamefont
  {Trave}}, \bibinfo {author} {\bibfnamefont {P.}~\bibnamefont {Tangney}},
  \bibinfo {author} {\bibfnamefont {S.}~\bibnamefont {Scandolo}}, \bibinfo
  {author} {\bibfnamefont {A.}~\bibnamefont {Pasquarello}}, \ and\ \bibinfo
  {author} {\bibfnamefont {R.}~\bibnamefont {Car}},\ }\bibfield  {title} {\emph
  {\bibinfo {title} {Pressure-induced structural changes in liquid {S}i{O}$_2$
  from ab initio simulations},\ }}\href {\doibase
  10.1103/PhysRevLett.89.245504} {\bibfield  {journal} {\bibinfo  {journal}
  {Phys. Rev. Lett.}\ }\textbf {\bibinfo {volume} {89}},\ \bibinfo {pages}
  {245504} (\bibinfo {year} {2002})}\BibitemShut {NoStop}%
\bibitem [{\citenamefont {Zeidler}\ \emph {et~al.}(2014)\citenamefont
  {Zeidler}, \citenamefont {Wezka}, \citenamefont {Rowlands}, \citenamefont
  {Whittaker}, \citenamefont {Salmon}, \citenamefont {Polidori}, \citenamefont
  {Drewitt}, \citenamefont {Klotz}, \citenamefont {Fischer}, \citenamefont
  {Wilding}, \citenamefont {Bull}, \citenamefont {Tucker},\ and\ \citenamefont
  {Wilson}}]{zeidler2014}%
  \BibitemOpen
  \bibfield  {author} {\bibinfo {author} {\bibfnamefont {A.}~\bibnamefont
  {Zeidler}}, \bibinfo {author} {\bibfnamefont {K.}~\bibnamefont {Wezka}},
  \bibinfo {author} {\bibfnamefont {R.~F.}\ \bibnamefont {Rowlands}}, \bibinfo
  {author} {\bibfnamefont {D.~A.}\ \bibnamefont {Whittaker}}, \bibinfo {author}
  {\bibfnamefont {P.~S.}\ \bibnamefont {Salmon}}, \bibinfo {author}
  {\bibfnamefont {A.}~\bibnamefont {Polidori}}, \bibinfo {author}
  {\bibfnamefont {J.~W.}\ \bibnamefont {Drewitt}}, \bibinfo {author}
  {\bibfnamefont {S.}~\bibnamefont {Klotz}}, \bibinfo {author} {\bibfnamefont
  {H.~E.}\ \bibnamefont {Fischer}}, \bibinfo {author} {\bibfnamefont {M.~C.}\
  \bibnamefont {Wilding}}, \bibinfo {author} {\bibfnamefont {C.~L.}\
  \bibnamefont {Bull}}, \bibinfo {author} {\bibfnamefont {M.~G.}\ \bibnamefont
  {Tucker}}, \ and\ \bibinfo {author} {\bibfnamefont {M.}~\bibnamefont
  {Wilson}},\ }\bibfield  {title} {\emph {\bibinfo {title} {High-pressure
  transformation of {S}i{O}$_2$ glass from a tetrahedral to an octahedral
  network: a joint approach using neutron diffraction and molecular dynamics},\
  }}\href {\doibase 10.1103/PhysRevLett.113.135501} {\bibfield  {journal}
  {\bibinfo  {journal} {Phys. Rev. Lett.}\ }\textbf {\bibinfo {volume} {113}},\
  \bibinfo {pages} {135501} (\bibinfo {year} {2014})}\BibitemShut {NoStop}%
\bibitem [{\citenamefont {Benmore}\ \emph {et~al.}(2010)\citenamefont
  {Benmore}, \citenamefont {Soignard}, \citenamefont {Amin}, \citenamefont
  {Guthrie}, \citenamefont {Shastri}, \citenamefont {Lee},\ and\ \citenamefont
  {Yarger}}]{benmore2010}%
  \BibitemOpen
  \bibfield  {author} {\bibinfo {author} {\bibfnamefont {C.}~\bibnamefont
  {Benmore}}, \bibinfo {author} {\bibfnamefont {E.}~\bibnamefont {Soignard}},
  \bibinfo {author} {\bibfnamefont {S.}~\bibnamefont {Amin}}, \bibinfo {author}
  {\bibfnamefont {M.}~\bibnamefont {Guthrie}}, \bibinfo {author} {\bibfnamefont
  {S.}~\bibnamefont {Shastri}}, \bibinfo {author} {\bibfnamefont
  {P.}~\bibnamefont {Lee}}, \ and\ \bibinfo {author} {\bibfnamefont
  {J.}~\bibnamefont {Yarger}},\ }\bibfield  {title} {\emph {\bibinfo {title}
  {Structural and topological changes in silica glass at pressure},\ }}\href
  {\doibase 10.1103/PhysRevB.81.054105} {\bibfield  {journal} {\bibinfo
  {journal} {Phys. Rev. B}\ }\textbf {\bibinfo {volume} {81}},\ \bibinfo
  {pages} {054105} (\bibinfo {year} {2010})}\BibitemShut {NoStop}%
\bibitem [{\citenamefont {Umari}\ \emph {et~al.}(2001)\citenamefont {Umari},
  \citenamefont {Pasquarello},\ and\ \citenamefont {Dal~Corso}}]{alf2001}%
  \BibitemOpen
  \bibfield  {author} {\bibinfo {author} {\bibfnamefont {P.}~\bibnamefont
  {Umari}}, \bibinfo {author} {\bibfnamefont {A.}~\bibnamefont {Pasquarello}},
  \ and\ \bibinfo {author} {\bibfnamefont {A.}~\bibnamefont {Dal~Corso}},\
  }\bibfield  {title} {\emph {\bibinfo {title} {Raman scattering intensities in
  $\alpha$-quartz: {A} first-principles investigation},\ }}\href {\doibase
  10.1103/PhysRevB.63.094305} {\bibfield  {journal} {\bibinfo  {journal} {Phys.
  Rev. B}\ }\textbf {\bibinfo {volume} {63}},\ \bibinfo {pages} {094305}
  (\bibinfo {year} {2001})}\BibitemShut {NoStop}%
\bibitem [{\citenamefont {Giacomazzi}\ \emph {et~al.}(2009)\citenamefont
  {Giacomazzi}, \citenamefont {Umari},\ and\ \citenamefont
  {Pasquarello}}]{giacomazzi2009}%
  \BibitemOpen
  \bibfield  {author} {\bibinfo {author} {\bibfnamefont {L.}~\bibnamefont
  {Giacomazzi}}, \bibinfo {author} {\bibfnamefont {P.}~\bibnamefont {Umari}}, \
  and\ \bibinfo {author} {\bibfnamefont {A.}~\bibnamefont {Pasquarello}},\
  }\bibfield  {title} {\emph {\bibinfo {title} {Medium-range structure of
  vitreous {S}i{O}$_2$ obtained through first-principles investigation of
  vibrational spectra},\ }}\href {\doibase 10.1103/PhysRevB.79.064202}
  {\bibfield  {journal} {\bibinfo  {journal} {Phys. Rev. B}\ }\textbf {\bibinfo
  {volume} {79}},\ \bibinfo {pages} {064202} (\bibinfo {year}
  {2009})}\BibitemShut {NoStop}%
\bibitem [{int()}]{intel_mkl}%
  \BibitemOpen
  \href@noop {} {}\bibinfo {note} {{I}ntel Corporation, Intel Math Kernel
  Library, https://software.intel.com/en-us/intel-mkl}\BibitemShut {NoStop}%
\bibitem [{\citenamefont {Bell}(1972)}]{bell72}%
  \BibitemOpen
  \bibfield  {author} {\bibinfo {author} {\bibfnamefont {R.}~\bibnamefont
  {Bell}},\ }\bibfield  {title} {\emph {\bibinfo {title} {The dynamics of
  disordered lattices},\ }}\href {\doibase 10.1088/0034-4885/35/3/306}
  {\bibfield  {journal} {\bibinfo  {journal} {Rep. Prog. Phys.}\ }\textbf
  {\bibinfo {volume} {35}},\ \bibinfo {pages} {1315} (\bibinfo {year}
  {1972})}\BibitemShut {NoStop}%
\bibitem [{\citenamefont {Pasquarello}\ \emph {et~al.}(1998)\citenamefont
  {Pasquarello}, \citenamefont {Sarnthein},\ and\ \citenamefont
  {Car}}]{alfprb98}%
  \BibitemOpen
  \bibfield  {author} {\bibinfo {author} {\bibfnamefont {A.}~\bibnamefont
  {Pasquarello}}, \bibinfo {author} {\bibfnamefont {J.}~\bibnamefont
  {Sarnthein}}, \ and\ \bibinfo {author} {\bibfnamefont {R.}~\bibnamefont
  {Car}},\ }\bibfield  {title} {\emph {\bibinfo {title} {Dynamic structure
  factor of vitreous silica from first principles: Comparison to
  neutron-inelastic-scattering experiments},\ }}\href {\doibase
  10.1103/PhysRevB.57.14133} {\bibfield  {journal} {\bibinfo  {journal} {Phys.
  Rev. B}\ }\textbf {\bibinfo {volume} {57}},\ \bibinfo {pages} {14133}
  (\bibinfo {year} {1998})}\BibitemShut {NoStop}%
\end{thebibliography}
%

%
\end{document}